\documentclass[fleqn,usenatbib]{mnras}

\usepackage{newtxtext,newtxmath}
\usepackage{xcolor}
\usepackage[T1]{fontenc}

\DeclareRobustCommand{\VAN}[3]{#2}
\let\VANthebibliography\thebibliography
\def\thebibliography{\DeclareRobustCommand{\VAN}[3]{##3}\VANthebibliography}

\usepackage{graphicx}	
\usepackage{amsmath}	
\usepackage{multirow}
\usepackage{bm}

\usepackage{algpseudocode}
\usepackage{algorithmicx}
\usepackage{algorithm}
\usepackage{empheq}
\usepackage{cases}

\newcommand{\xb}{\ensuremath{\boldsymbol{x}}}
\newcommand{\yb}{\ensuremath{\boldsymbol{y}}}

\newcommand{\nb}{\ensuremath{\boldsymbol{n}}}

\newcommand{\Gb}{\ensuremath{\boldsymbol{\mathsf{G}}}}

\newcommand{\Hb}{\ensuremath{\boldsymbol{\mathsf{H}}}}
\newcommand{\Wb}{\ensuremath{\boldsymbol{\mathsf{W}}}}

\newcommand{\eC}{\mathbb{C}}
\newcommand{\eR}{\mathbb{R}}
\newcommand{\eN}{\mathbb{N}}

\title[uSARA-ASKAP]{Scalable precision wide-field imaging in radio interferometry: \\ I. uSARA validated on ASKAP data}

\author[A. G. Wilber et al.]{
A. G. Wilber,$^{1}$
A. Dabbech,$^{1}$
A. Jackson,$^{2}$
Y. Wiaux$^{1}$\thanks{E-mail: y.wiaux@hw.ac.uk}
\\
$^{1}$Institute of Sensors, Signals and Systems, Heriot-Watt University, Edinburgh EH14 4AS, UK\\
$^{2}$EPCC, The University of Edinburgh, Edinburgh EH8 9BT, UK\\
}

\date{Accepted XXX. Received YYY; in original form ZZZ}

\pubyear{2023}

\begin{document}
\label{firstpage}
\pagerange{\pageref{firstpage}--\pageref{lastpage}}
\maketitle

\begin{abstract}
As Part I of a paper series showcasing a new imaging framework, we consider the recently proposed unconstrained Sparsity Averaging Reweighted Analysis (uSARA) optimisation algorithm for wide-field, high-resolution, high-dynamic range, monochromatic intensity imaging. We reconstruct images from real radio-interferometric observations obtained with the Australian Square Kilometre Array Pathfinder (ASKAP) and present these results in comparison to the widely-used, state-of-the-art imager {\tt WSClean}. Selected fields come from the ASKAP Early Science and Evolutionary Map of the Universe (EMU) Pilot surveys and contain several complex radio sources: the merging cluster system Abell 3391-95, the merging cluster SPT-CL 2023-5535, and many extended, or bent-tail, radio galaxies, including the X-shaped radio galaxy PKS 2014-558 and ``the dancing ghosts,'' known collectively as PKS 2130–538. The modern framework behind uSARA utilises parallelisation and automation to solve for the $w$-effect and efficiently compute the measurement operator, allowing for wide-field reconstruction over the full field-of-view of individual ASKAP beams (up to $\sim 3.3 ^{\circ}$ each). The precision capability of uSARA produces images with both super-resolution and enhanced sensitivity to diffuse components, surpassing traditional CLEAN algorithms which typically require a compromise between such yields. Our resulting monochromatic uSARA-ASKAP images of the selected data highlight both extended, diffuse emission and compact, filamentary emission at very high resolution (up to 2.2 arcsec), revealing never-before-seen structure. Here we present a validation of our uSARA-ASKAP images by comparing the morphology of reconstructed sources, measurements of diffuse flux, and spectral index maps with those obtained from images made with {\tt WSClean}.      
\end{abstract}

\begin{keywords}
techniques: interferometric -- techniques: image processing -- radio continuum: galaxies -- galaxies: clusters: intracluster medium 
\end{keywords}



\section{Introduction}
\label{sec:intro}

We are now at the beginning of a booming era for radio astronomy, with a worldwide effort to gear up for the Square Kilometre Array (SKA) -- a revolutionary radio telescope with capabilities for sub-arcsecond resolution and ultra-deep sensitivity. Pathfinding radio interferometers -- such as the Murchison wide-field Array (MWA; \citealp{2013PASA...30....7T}), the LOw Frequency ARay (LOFAR; \citealp{2013A&A...556A...2V}), ASKAP \citep{2007PASA...24..174J, 2008ExA....22..151J, 2021PASA...38....9H}, and MeerKAT \citep{2016mks..confE...1J} -- are paving the way by affording us the opportunity to expand our capabilities in detection, calibration, and image reconstruction of the unknown radio sky. Ongoing wide-field radio continuum surveys -- such as the LOFAR Two-meter Sky Survey (LoTSS; \citealp{2017A&A...598A.104S,2019A&A...622A...1S}), the LOFAR LBA Sky Survey (LoLSS; \citealp{2021A&A...648A.104D}), the MeerKAT MIGHTEE survey \citep{Taylor_2017}, and the ASKAP EMU survey \citep{2011PASA...28..215N} -- will be used in conjunction to gather statistics on millions of radio galaxies and thousands of galaxy clusters, leading to new and exciting results on cosmic magnetism, cosmic rays, dark matter, dark energy, and the evolution of large-scale structure in the Universe. The quest to convert the sheer quantity of radio data from these surveys into science-ready images has propelled radio astronomers toward developing innovative, state-of-the-art calibration \citep[\emph{e.g.}][]{2015MNRAS.449.2668S,2016MNRAS.460.2385W,2016ApJS..223....2V} and imaging \citep[\emph{e.g.}][]{2017MNRAS.471..301O,2018A&A...611A..87T, 2018MNRAS.473.1038P} techniques throughout the last decade. Through the implementation of these trailblazing techniques, results from LOTSS, the ASKAP-EMU Pilot Survey \citep[][]{2021PASA...38...46N}, and MIGHTEE are already revealing extraordinary extragalactic radio sources that have not been previously detected, some of which are so complex in their physical properties that they challenge current taxonomies and origination theories \citep[\emph{e.g.}][]{2020ApJ...897...93B, 2021A&A...647A...3B, 2022A&A...657A..56K}.

Real intensity structure in radio sources -- such as Galactic HI emission, supernova remnants, extended radio galaxies, and galaxy cluster radio halos and relics -- often exhibits both compact and diffuse components. A complex radio source might include bright, collimated threads of emission that appear embedded within fainter, dispersed lobes. For instance, intracluster radio sources, generated by large-scale turbulence and shocks, tend to exhibit high-resolution filamentary structures (tracing well-aligned magnetic field lines) as well as low-surface-brightness diffuse structure, which can span Mega-parsec scales (see the radio relic of Abell 2256 as one iconic example --  \citealp[\emph{e.g.}][]{2014ApJ...794...24O,2022ApJ...927...80R}). When imaging such complex radio emission, state-of-the-art CLEAN-based algorithms \citep[\emph{e.g.}][]{1974A&AS...15..417H,1980A&A....89..377C,1984AJ.....89.1076S,1988A&A...200..312W,2008ISTSP...2..793C, 2014MNRAS.444..606O} are usually implemented with data weighting schemes to adjust the sensitivity to either compact or diffuse emission. By modifying the shape of the synthesised beam and placing weights (such as a $uv$-taper) to short-baseline data, the sensitivity to fainter, more extended components of radio emission can be enhanced, albeit with a considerable loss of resolution. Consequently, such imaging methods often fail to accurately reconstruct both compact and diffuse components simultaneously in a single image.

In the last decade, progress in compressed sensing research has led to the development of optimisation-based algorithms to reconstruct true signal from partial, undersampled visibility data. This methodology was first applied and shown to be effective for radio interferometry imaging by \citet{2009MNRAS.395.1733W} and has since led to the development of several designated imaging algorithms \citep[e.g.][]{2011A&A...528A..31L, Carrillo12,Garsden2015,2015A&A...576A...7D}. Such innovative approaches -- relying on sophisticated sparsity-based image models -- have demonstrated their success at capturing both compact and diffuse components in the reconstructed radio images, albeit with an increase in computational expense. One such state-of-the-art optimisation-based class of methods is the ``Sparsity Averaging Reweighted Analysis'' (SARA) family \citep[][see the \href{https://basp-group.github.io/Puri-Psi/}{Puri-Psi} web-page for more details]{Carrillo12,Dabbech18,Abdulaziz19,thouvenin22a}. The monochromatic SARA algorithm, initially proposed by \citet{Carrillo12}, leverages state-of-the-art algorithmic structures from optimisation theory to enforce non-negativity and sparsity in a highly redundant sparsity dictionary
of the sought radio image under noise-driven data fidelity constraints \citep{onose2016,onose17}. Evolved versions of SARA include Faceted HyperSARA for wide-band imaging, shipped with spectral and spatial faceting functionalities to handle large image dimensions \citep{thouvenin22a,thouvenin22b}, Polarized SARA for polarisation imaging \citep{birdi18}, and a sparsity-based joint calibration and imaging framework \citep{repetti17,repettiProc17,birdi20,Dabbech21}.

In addition to the precision and robustness requirements of RI imaging algorithms, the need for scalability is more critical than ever in light of the extreme data volumes, wide fields-of-view, and broad frequency bandwidths offered by modern radio arrays. In this context, we have recently proposed a parallel and automated framework for wide-field, high-resolution, high-dynamic range monochromatic intensity imaging \citep{Dabbech22}. The framework encapsulates two imaging algorithms at the interface of optimisation theory and deep learning, recently proposed by \citet{Terris22}: (i) the unconstrained SARA (uSARA) algorithm relying on a handcrafted image model enforced via an optimisation-based denoiser, and (ii) the AI for Regularization in Imaging (AIRI) algorithm relying on an implicitly learnt image model promoted via denoising deep neural networks. The framework offers scalable implementation of both imaging algorithms through two key features:  memory-efficient, parallel operators and automated parameter selection.

This article is Part I of a series aiming to showcase and validate the uSARA algorithm on imperfectly calibrated RI observations from the ASKAP radio telescope. In Part II, we expand upon this work to include a validation of the AIRI algorithm. In both articles, we aim to study the imaging performance of uSARA and AIRI in comparison to Multi-scale CLEAN \citep{2008ISTSP...2..793C} via the {\tt WSClean} imager \citep{2014MNRAS.444..606O}, both in terms of reconstruction quality and computational efficiency. For a coherent comparative analysis of the three imaging algorithms --  uSARA, AIRI, and {\tt WSClean} --  both Part I and Part II utilise the same RI data from publicly available ASKAP observations collected during Early Science and Pilot surveys and processed through the ASKAPsoft pipeline \citep{2021PASA...38....9H}. Targeted fields-of-view -- hosting extended, diffuse, and complex radio sources -- were carefully selected to test the precision and scalability capabilities of our imaging framework. A comprehensive summary of the considered imaging framework (with interchangeable denoising algorithms), including the wide-field measurement operator model, its distribution through parallelisation, and its implementation using high-performance computing systems (HPC), is provided by \citet{Dabbech22}. A fully detailed analysis of the framework's scalability is the subject of a forthcoming article. 

The remainder of this article is structured as follows. In Section~\ref{sec:methods}, we present an overview of the investigated imaging framework from the algorithmic structure underpinning uSARA to the parallelisation and automation functionalities ensuring the computational scalability of the framework to large image and data dimensions in the context of wide-field imaging. In Section~\ref{sec:data}, we provide details of the scrutinised ASKAP data and the imaging settings of uSARA and the CLEAN-based benchmark algorithm. Reconstruction results of our primary targets of interest are exposed in Section~\ref{sec:results} and discussed in Section~\ref{sec:disc}. Section~\ref{sec:comp-cost} documents and discusses the computational cost of the imaging framework.  Finally, conclusions are drawn in Section~\ref{sec:con}.

\section{Methods}\label{sec:methods}

In this section, we present the RI data model in the context of wide-field monochromatic intensity imaging. We provide an overview of the uSARA imaging algorithm and its underpinning algorithmic structure, and summarise the scalability features of its encompassing framework \citep{Terris22,Dabbech22}.

\subsection{Data Model}

\label{ssec:data}
In the absence of instrumental and atmospheric perturbations, RI visibilities measured at a given observing wavelength are noisy Fourier components of the radio sky, modulated by the so-called $w$-effect, a varying chirp-like phase induced by the non-coplanarity of the radio array. With no loss of generality, the data model can be discretised, such that the measured visibilities $\yb\in \eC^M$ are modelled from the sought intensity image ${\xb}\in\eR^N_+$ as follows
\begin{equation}
    \label{eq:invpb}
     \yb=\bm{\Phi} {\xb}+{\nb},
\end{equation}
where $\nb \in \eC^M$ is a realisation of a zero-mean random Gaussian noise with a standard deviation $\tau >0$. The operator $\bm{\Phi}\in \eC^{M\times N}$ is the measurement operator encompassing the Fourier sampling, and the $w$-effect. {Due to the large amounts of data, a Direct Fourier transform would be intractable; therefore, the incomplete Fourier sampling is} modelled via the non-uniform fast Fourier transform (NUFFT) \citep{Fessler2003,onose2016}. {Furthermore,} the $w$-effect is taken into account via a hybrid model combining $w$-stacking \citep{2014MNRAS.444..606O} and $w$-projection \citep{Cornwell2008}, whereby RI data are grouped by their $w$-coordinates into $P$ $w$-stacks. For each data point, the chirp-like phase is decomposed into two terms: a modulation of its associated $w$-stack injected in the measurement operator via image-domain multiplication, and a compact Fourier kernel encoding the resulting $w$-offset modulation, injected via Fourier-domain convolution \citep{dabbech17}. As a final consideration, a noise-whitening operation is typically applied to the measured data and injected into the associated measurement operator to ensure constant standard deviation of the noise \citep[see Appendix A of ][for more details]{Terris22}. On some occasions, the operation is performed in combination with a data-weighting scheme derived from the sampling profile (\emph{e.g.}~ Briggs weighting; \citealp{briggs95}) to improve the effective resolution of the observation (due to the highly non-uniform density profile of the RI sampling). 

Under these considerations, the measurement operator is decomposed into computationally and memory-efficient blocks as the vertical concatenation of the operators $\big(\bm{\Phi}_p\big)_{1\leq p\leq P}$, where for each $w$-stack $p\in\{1,\dots,P\}$, the associated measurement operator $\bm{\Phi}_p \in \eC^{M_p \times N} $ is given by $\bm{\Phi}_p= \bm{\Theta}_p{\mathbf{G}}_p {\mathbf{F}} {\mathbf{Z}}_p$ \citep{Dabbech22}. More specifically, the operator $\bm{\Theta}_p\in \eR^{M_p \times M_p}$ is a diagonal matrix encoding the considered data-weighting scheme. The sparse matrix $\mathbf{G}_p\in \eC^{M_p \times N^\prime}$ is the de-gridding matrix, encompassing convolutions between the NUFFT interpolation kernels and the compact $w$-kernels correcting for the associated $w$-offsets in the Fourier plane. Note that estimates of direction-dependent effects (DDEs) can also be encoded as additional convolutions in the rows of the de-gridding matrix, when available. $\mathbf{F} \in \eC^{N^\prime \times N^\prime}$ is the Discrete Fourier transform and the operator $\mathbf{Z}_p \in \eC^{N^\prime \times N}$ encodes the $w$-modulation of the $p^{\textrm{th}}$ $w$-stack, the zero-padding operator for a finer grid of the Fourier plane, and the correction for the convolution with the approximate NUFFT interpolation kernels. 

\subsection{uSARA algorithm}

Image formation from the noisy and incomplete RI measurements $\yb$ is an ill-posed inverse problem. Here we consider the unconstrained SARA imaging algorithm from optimisation theory \citep{Terris22}. The algorithm provides an estimate of the radio sky as the minimiser of an objective function posed as the sum of two terms: a data fidelity term $f$, emanating from the nature of the noise, and a regularisation term $r$ encoding a prior knowledge of the image to address the ill-posedness of the inverse problem. The minimisation task is of the form
\begin{equation}
\label{eq:objfun}
 \underset{\xb\in \eR^N}{\mathrm{minimise}} ~~ f(\xb; \yb)+ \lambda r(\xb),
\end{equation}
where $\lambda>0$ is the regularisation parameter controlling the balance between the two terms. 
Given the Gaussian nature of the noise affecting the RI data, $f$ is naturally set to $f(\xb; \yb)=1/2\|\bm{\Phi}\xb-\yb\|_2^2$, with $\| .\|_2$ denoting the $\ell_2$ norm of its argument vector.

The uSARA regularisation function $r$ is a multi-term non-differentiable function composed of a non-convex log-sum function enforcing average sparsity in an overcomplete dictionary $\bm{\Psi} \in \eR^{N\times B}$, consisting in the normalised concatenation of nine orthogonal bases, and a non-negativity constraint \citep{Carrillo12,Terris22}, which reads
\begin{equation}
    r(\xb) = \rho \sum_{j=1}^{B} \log\left(\rho^{-1}\left|\left(\bm{\Psi}^\dagger \xb\right)_{j}\right|+1\right)+\iota_{\eR^N_+}(\xb),
\label{eq:sara_prior}
\end{equation}
where $\left(.\right)_j$ denotes the $j^\text{th}$ coefficient of its argument vector, and  $(\cdot)^\dagger$ stands for the adjoint of its argument operator. The parameter $\rho>0$ prevents the argument of the logarithmic from reaching zero values and can be set to the estimate of the noise level in the sparsity domain \citep{thouvenin22a}. The non-negativity constraint is encoded via the indicator function of the real positive orthant, given by $\iota_{{\eR^N_{+}}} (\xb)=+\infty$ if $\xb \notin {\eR^N_{+}}$ and 0 otherwise. As such, the resulting minimisation task is non-convex and is addressed in an iterative manner. More specifically, the problem is approached by solving a sequence of surrogate convex minimisation tasks whereby $r$ is replaced by a convex regularisation function $g$ of the form $g(\xb)=  \|\Wb \bm{\Psi}^\dagger \xb\|_1+\iota_{\eR^N_+}(\xb)$ substituting the log-sum function with the $\ell_1$ function, denoted by $\|\cdot\|_1$, and weighted by the diagonal matrix $\Wb \in\eR^{B \times B}$. Each of the surrogate weighted minimisation tasks is the form 
\begin{equation}
\label{eq:objfun2}
 \underset{\xb\in \eR^N}{\mathrm{minimise}} ~~ f(\xb; \yb)+ \lambda g(\xb),
\end{equation}
where the convex and non-differentiable function $g$ is redefined through the update of its underlying weighting matrix $\Wb$ from the solution of its preceding task \citep{Carrillo12,Terris22}. The convex minimisation task is solved approximately (\emph{i.e.} for a finite number of iterations $K>0$) using the forward-backward (FB) iterative scheme \citep{repetti20,Terris22}, and the overall procedure benefits from convergence guarantees \citep{repetti21}.

The FB iterative scheme relies on two-step image updates: a `forward' gradient descent step calling for the gradient of the data fidelity function $f$, given by $\nabla f(\xb )= \text{Re}\{\bm{\Phi}^\dagger\bm{\Phi}\} \xb -\text{Re}\{\bm{\Phi}^\dagger\yb\}$, followed by a `backward' denoising step using the proximal operator of the convex regularisation function $g$ \citep[see ][for the mathematical details]{Terris22} such that for any $k \in \eN$
\begin{equation}
\label{eq:fb}
 \xb^{(k+1)} = \operatorname{prox}_{\gamma \lambda g}\left( \xb^{(k)} -\gamma \nabla f(\xb^{(k)}) \right).
\end{equation}
Let $L>0$ denote the Lipschitz constant of $\nabla f$ given by $L=\| \text{Re}\{\bm{\Phi}^\dagger \bm{\Phi}\}\|_S$, with $\| .\|_S$ denoting the spectral norm of its argument operator. The step-size $\gamma$ satisfies the condition $0<\gamma<2/L$ to guarantee the convergence of the iterative scheme. Finally, the proximal operator $\operatorname{prox}_{\gamma \lambda g}$, not benefiting from closed-form solutions, is computed sub-iteratively, involving soft-thresholding operations in the sparsity dictionary $\bm{\Psi}$  by $\gamma \lambda$.

\subsection{A scalable and automated imaging framework}\label{sec:regparam}

To address the scalability requirement to large data and image sizes, our imaging framework provides automated parameter choice and parallel and memory-efficient models of the operators and functions involved~\citep{Dabbech22}, summarised in what follows. 

\paragraph*{Regularisation parameter selection.} 

The choice of the regularisation parameter $\lambda$, balancing data fidelity and image regularisation, is of paramount importance as it affects the solution of the minimisation task \eqref{eq:objfun}. Considering $\sigma>0$, the estimate of the standard deviation of the image domain noise, \citet{Terris22} proposed to equate the soft-thresholding parameter $\gamma \lambda$, involved in the denoising step, to the estimate of the standard deviation of the sparsity domain noise given by $\sigma/3 $ (the factor three emanates from the normalisation of the sparsity dictionary). In the case when a data-weighting scheme is adopted to compensate for the non-uniform density profile of the sampling (\emph{e.g.} Briggs weighting), additional correlation is induced in the image domain noise. Under this consideration, $\sigma$ can be obtained as
\begin{equation}
\label{eq:sigma}
    \sigma=\eta \tau/\sqrt{2L}, \textrm{~with~}  \eta={\| \text{Re}\{\bm{\Phi}^\dagger \bm{\Theta}^2\bm{\Phi}\}\|^{1/2}_S}/{\sqrt{L}},
\end{equation}
where the data-weighting operator $\bm{\Theta}\in \eR^{M\times M}$ is a diagonal per block matrix, whose diagonal matrix-blocks are the data-weighting matrices $\big(\bm{\Theta}_p\big)_{1\leq p\leq P}$. The correction factor $\eta$ reduces to one otherwise. In our experiments, Briggs weighting was applied to the data in imaging, and the resulting values of $\eta$ were found to be in the interval $[0.3,0.6]$. The regularisation parameter $\lambda$ can be set around 
\begin{equation}
    \label{eq:heuristic}
    \lambda \simeq \tau {\| \text{Re}\{\bm{\Phi}^\dagger \bm{\Theta}^2\bm{\Phi}\}\|^{1/2}_S} /(3\sqrt{2}\gamma L ),
\end{equation}
with the step size fixed to $\gamma=1.98/L$, ensuring the convergence of the FB algorithm. {Finally, the parameter $\rho$ involved in \eqref{eq:sara_prior} is typically set to the estimate of the standard deviation of the sparsity domain noise, $\rho=\sigma/3$.}

\paragraph*{Denoiser Faceting.}

In light of the sub-iterative nature of the denoising operator underpinning uSARA, distribution and parallelisation of the sparsity operator $\bm{\Psi}$ are required, not only to handle the large image dimensions of interest but also to accelerate the computation. For this aim, we have adopted a faceted implementation of the operator $\bm{\Psi}$ \citep{Prusa2012}, enabling image facet denoising. The number of facets $F$ is derived from the number of CPU cores of the computing architecture on which the algorithm is to be deployed, and constraints on image facet dimensions from the wavelet transforms underpinning the sparsity operator $\bm{\Psi}$.

\paragraph*{Automated parallelisation of the measurement operator.} 
\label{ssec:prallel} Three key features are supported in our implementation of the measurement operator to ensure its scalability to large data sizes \citep{Dabbech22}.
Firstly, the choice of the number of $w$-stacks, $P$, defining the decomposition of the measurement operator into the operators $\bm{\Phi}_p$ is automated via a planning strategy taking into consideration the computational cost derived from the complexity of the application of the measurement operator $\bm{\Phi}$ and the memory constraints of the computing architecture. Secondly, memory-efficient encoding of the resulting operators $\bm{\Phi}^\dagger_p \bm{\Phi}_p$, called for in FB, can be achieved through a data dimensionality reduction functionality via visibility-gridding, whereby the de-gridding and gridding operations underpinning  $\bm{\Phi}^\dagger_p \bm{\Phi}_p$ are explicitly encoded via the sparse holographic matrices $\Hb_p ={\Gb}^\dagger_p \bm{\Theta}^2_p \Gb_p$. By doing so, the dimensionality of the measurement operator is effectively driven solely by the image size. The feature is enabled when the memory required to host the de-gridding matrices exceeds the available resources. Thirdly, further decomposition of each operator $\bm{\Phi}_p^\dagger\bm{\Phi}_p$ into smaller blocks is enabled via a data-clustering step such that each data block corresponds to the aggregation of radially-neighbouring Fourier modes, identified under memory constraints. The number of CPU cores allocated for the forward step of the FB algorithmic structure is derived from the number of identified data clusters. These computing resources are initially used to compute the de-gridding matrices ($\Gb_{p})$ or the holographic matrices ($\Hb_{p})$, underpinning the operator $\bm\Phi^\dagger \bm\Phi$ (only once), and are later used to host them and apply them at each FB iteration.

\begin{table*}
\centering
\begin{center}
\begin{tabular}{ccccccc} 
\textbf{SB -- Beam} & \textbf{Band} & \textbf{Sources} & \textbf{R.A., Dec.} & \textbf{Redshift}  & \textbf{Features} \\
 & [MHz] &  & (J2000) &   & \\
\hline
\multirow{2}{4em}{8275--15} & \multirow{2}{5em}{870 - 1158} & Abell 3391 & 06h26m22.8s, $-53^{\circ}41'44''$  & $z=0.056^{(1)}$ & North MC, FR I  \\ 
&  & Abell 3395 & 06h27m14.4s, $-54^{\circ}28'12''$ & $z=0.0518^{(1)}$ & South MC, FR I, radio phoenix\\ 
\hline
\multirow{2}{4em}{9351--12} & \multirow{2}{5em}{800 - 1088} & PKS B2014-558 & 20h18m01.1s, $-55^{\circ}39'31''$  & $z = 0.061^{(2)}$ & X-shaped RG \\ 
&  & SPT-CL J2023-5535 & 20h23m24.5s, $-55^{\circ}35'32''$ & $z=0.23^{(3)}$ & radio halo, radio relic \\ 
\hline
\multirow{2}{4em}{9442--35} & \multirow{2}{5em}{800 - 1088} & Abell 3785 &  \multirow{2}{10em}{21h34m28s, $-53^{\circ}37'$}  & \multirow{2}{6em}{$z = 0.0773^{(4)}$} & complex RGs \\ 
&  & PKS 2130-538 &  &  & ``the dancing ghosts'' \\
\hline
\end{tabular}
\caption{Selected observations from the ASKAP Early Science and EMU Pilot Surveys for imaging. Datasets are labelled by their ASKAP Scheduling Block (SB) and primary beam pointing ID. Multiple sources of interest are covered per dataset, as listed, with notes in the `Features' column. Redshifts are from (1) \citealp[][]{2004AJ....128.1558S} (2) \citealp[][]{Huchra_2012} (3) \citealp[][]{2021ApJS..253....3H} (4) \citealp[][]{1999ApJS..125...35S}. MC: Merging cluster. FR I: Fanaroff-Riley class I. RG: Radio galaxy.} \label{tab: obs}
\end{center}
\end{table*}

\begin{table}
    \begin{tabular}{ccc}
        \textbf{Parameters} & Sub-band Images & Full-band Image$^{~\star\star}$\\
         \hline
        Number of SPWs & 8 & 1 \\
        Bandwidth & 36 MHz & 288 MHz\\
        \hline
        Field of View & $3.36^{\circ^{~\star}}$ & $2.5^{\circ}$\\
        & $2.5^{\circ~^{\star\star}}$ & \\
        \hline
        Image Size &  $5500 \times 5500^{~\star}$ & $4096 \times 4096$\\
        &  $4096 \times 4096^{~\star\star}$ & \\
        \hline
        Cell Size & \multicolumn{2}{c}{$2.2$ arcsec pixel$^{-1}$} \\
        \hline
        $uv$-range & \multicolumn{2}{c}{$>60$ m}  \\
        \hline
        Data weighting & \multicolumn{2}{c}{Briggs Robust $-0.25$} \\
        \hline
    \end{tabular}
     \caption{Imaging settings of uSARA and {\tt WSClean}. The notation ($^{\star}$) refers to the settings of the fields SB8275-15 and SB9351-12 and ($^{\star\star}$) refers to the settings of the field SB9442-35. \label{tab: imgparams}}
\end{table}

\section{Data, imaging, and analysis}\label{sec:data}

In this section, we provide a full description of the scrutinised data, including pre-processing and calibration steps. We provide the imaging settings of uSARA and {\tt WSClean}, and outline the computing architecture and resources required to run both algorithms. Finally, we provide procedures for a coherent comparative analysis. 

\subsection{ASKAP Observations}

ASKAP consists of 36 12-metre parabolic dish antennas, spanning six kilometres, at the Murchison Radio Observatory in Western Australia. ASKAP's original design includes Phased Array Feeds (PAFs; \citealp{2006ESASP.626E.663H}) at the focus of each antenna, built as dual-polarisation chequerboard grids sensing signals in the frequency range of 700-1800 MHz. Signals received by each of the 188-element-sensor grids are cross-correlated to simultaneously form 36 separate primary beams (or pointings) on the sky \citep{2014PASA...31...41H, 2016PASA...33...42M}. This PAF technology gives ASKAP an instantaneous field-of-view (FoV) of 30-square-degrees, making it the most rapid surveying radio telescope in the world \citep{2009IEEEP..97.1507D}. ASKAP's EMU Survey will survey the radio continuum over the entire southern sky (up to a northern declination of +30$^{\circ}$) at a resolution of $\sim 10$ arcsec with sensitivities reaching $\sim 10 \mu$ Jy beam$^{-1}$, and is projected to detect more than 70 million galaxies. Science goals of the EMU collaboration include testing fundamental models for dark energy \citep[\emph{e.g.}][]{2012MNRAS.424..801R}, detecting the warm-hot intergalactic medium \citep[\emph{e.g.}][]{2020PASA...37...32H}, tracing active galactic nuclei (AGN) and star formation up to high-redshifts \citep[\emph{e.g.}][]{2017ApJ...842...95M}, and mapping the radio continuum of the Galactic Plane and Centre \citep[\emph{e.g.}][]{2021MNRAS.502...60R}. 

The ASKAP Early Science Broadband Survey (Project: AS034, PI: Lisa Harvey-Smith) began in 2018 with the aim to test observations using the full ASKAP array while covering scientifically interesting areas of sky, such as the Galaxy And Mass Assembly 23-hour field (GAMA 23). The EMU Pilot Survey \citep[EMU-PS;][Project: AS101, PI: Ray Norris]{2021PASA...38...46N}, centred on RA, Dec: 21h00m00s, -55$^{\circ}$00$'$00$''$, was carried out in 2019  covering an area of 270 deg$^2$ overlapping a portion of sky covered by the first data release of the Dark Energy Survey (DES; \citealp{2018ApJS..239...18A}). From these early ASKAP data releases, we have selected three individual Scheduling Block (SB) beam observations covering extended, morphologically complex radio sources which represent robust test cases for image reconstruction with uSARA. ASKAP SBs contain 36 measurement sets each, corresponding to the 36 primary beam pointings. Therefore, to ensure maximum signal-to-noise, we have chosen the single beam measurement sets which were most closely centred on our primary targets of interest (beam 15 of SB8275, beam 12 of SB9351, and beam 35 of SB9442). Beam observations for each SB are carried out with the same specifications over a 10-hour total integration time. The observing band varies slightly between the selected Early Science data (SB8275 - central frequency 1013 MHz) and EMU-PS data (SB9351 and SB9442 - central frequency 943 MHz), yet both have instantaneous bandwidths of 288 MHz with physical channel intervals of 1 MHz. See Table~\ref{tab: obs} for further details on the selected observations.

The three selected SB-beam measurement sets, containing calibrated visibilities, were generated through the ASKAPsoft pipeline \citep{2021PASA...38....9H}. For direction-independent calibration, ASKAPsoft includes bandpass calibration using the standard calibrator PKS B1934-638 -- observed for five minutes in each beam before or after the science target. Further calibration includes a cycle of phase-only self-calibration for every 1 MHz of data. Each beam observation in an SB is calibrated and imaged independently, and as a final step, ASKAPsoft stitches images together to form mosaics covering the full 30-square-degree FoV. For our imaging purposes, we used the ASKAP data products after ASKAPsoft processing, with no further flagging or calibration. Prior to imaging, we shifted the pointing centres of our selected beam observations by their beam offsets. Although most RI packages assume a Stokes parameter $I$ (intensity) of $I = (XX + YY)/2$ (where $X$ and $Y$ represent the instrumental polarisations), ASKAPsoft uses the IAU (International Astronomical Union) definition $I = XX + YY$. We did not apply a factor two correction for the IAU intensity convention in any of our final image reconstructions; therefore, flux densities in our images are halved when compared to the values found in ASKAPsoft mosaic images.

\begin{figure*}
\centering
\includegraphics[width=\textwidth]{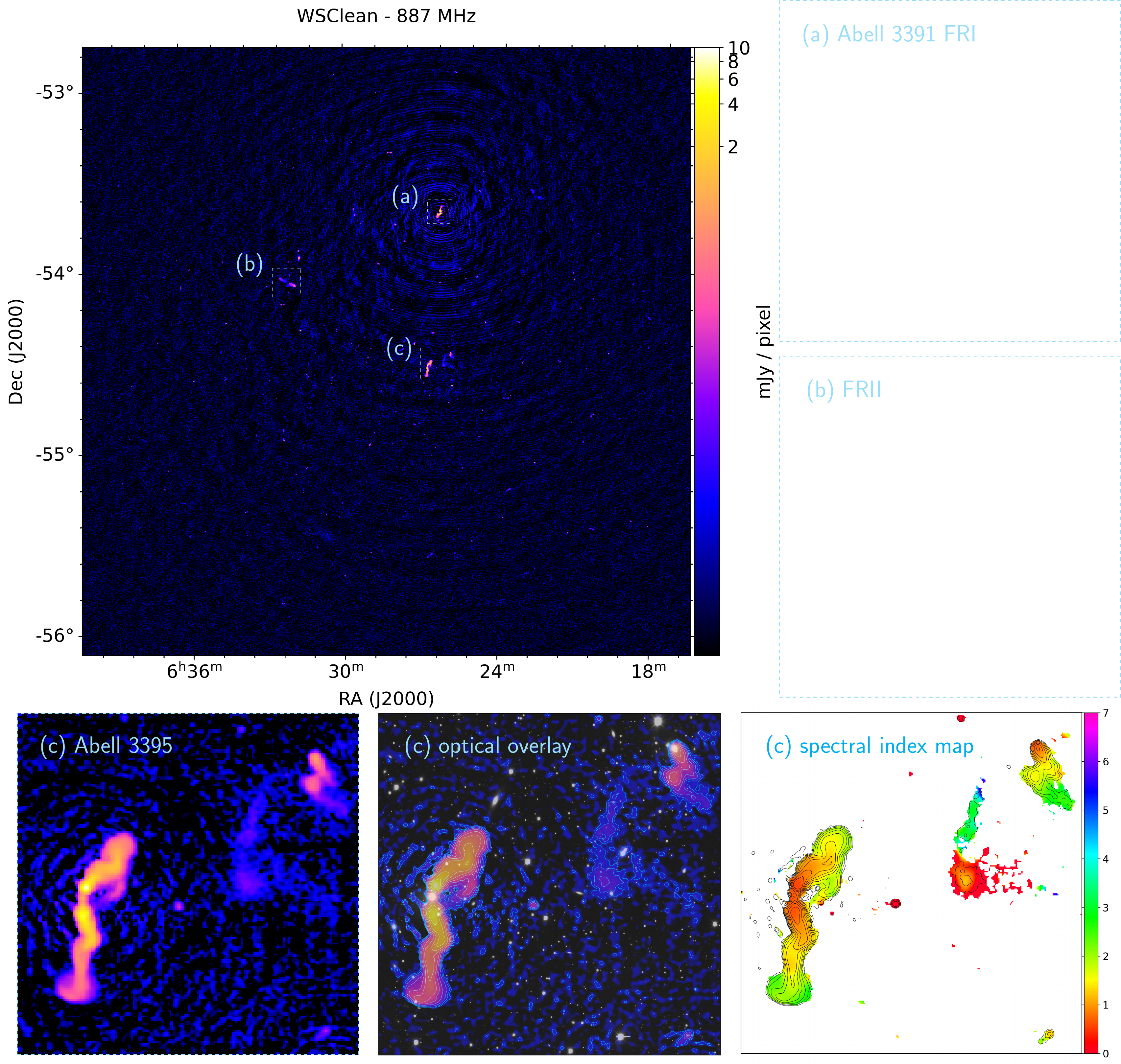}
\caption{SB8275-15 -- {\tt WSClean}: Full FoV image covering the merging cluster system Abell 3391-95, at the first sub-band (SPW:1, centred at 887 MHz). This monochromatic image is a {\tt WSClean} restored image with a synthesised beam of $9.4 \times 10.9$ arcsec and rms noise of $\sigma_{\rm meas}\approx 50~\mu$Jy beam$^{-1}$ ($2~\mu$Jy pixel$^{-1}$). Panel (a) centred on the FR I radio galaxy in A3391; panel (b) centred on cluster member FR II radio galaxy; (c) panels centred on FR I and diffuse source in A3395. Middle (c) panel: r-band optical image from DES overlaid with {\tt WSClean} restored image, demarcated
by blue contours at levels $\{2^{n+1}\}_{1 \leq n \leq 10} ~\mu$Jy pixel$^{-1}$. Rightmost (c) panel: Spectral index map obtained with the first six sub-band images of {\tt WSClean} after smoothing with a common circular Gaussian beam of 20 arcsec. In \citet{askapdataset} are provided all sub-band images combined into the GIF \texttt{`SB8275-15\_WSClean'}, and {the spectral index maps of Abell 3395 obtained with {\tt WSClean} and uSARA  in the GIF \texttt{`SpectralIndexMap\_Abell\_$3395$'}, together with a colour blind-friendly version in the GIF \texttt{`SpectralIndexMap\_Abell\_3395\_colorblind\_friendly'}.}
\label{A3391wsclean}}
\end{figure*}

\begin{figure*}
\centering
\includegraphics[width=\textwidth]{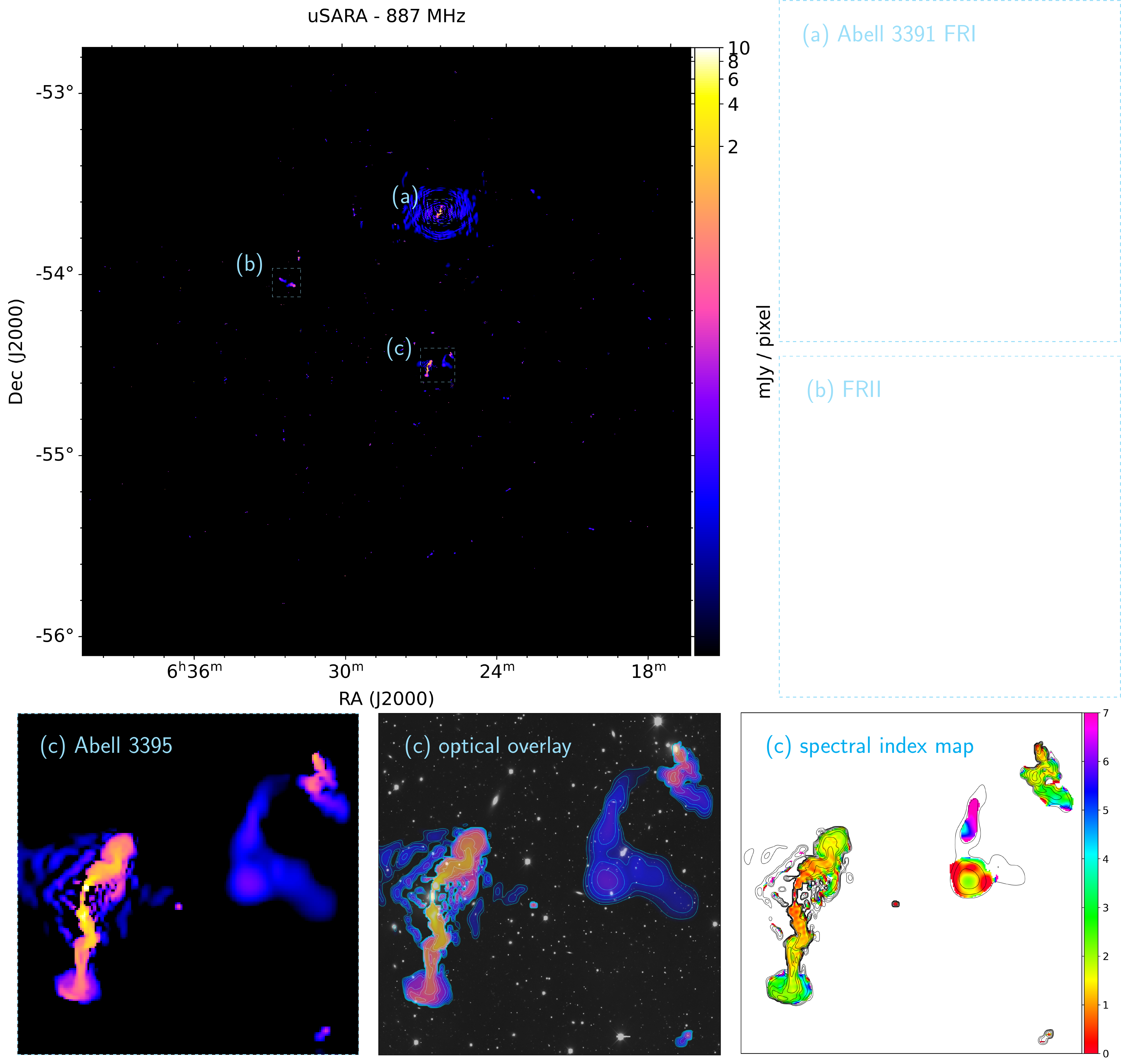}
\caption{SB8275-15 -- uSARA: Full FoV image covering the merging cluster system Abell 3391-95, at the first sub-band (SPW:1, centred at 887 MHz). This monochromatic image is a uSARA model with a pixel resolution of $2.2 \times 2.2$ arcsec. Panels are the same as described in Figure~\ref{A3391wsclean}. Middle (c) panel: r-band optical image from DES overlaid with uSARA model image, demarcated by blue contours at levels $\{2^n\}_{0 \leq n \leq 10} ~\mu$Jy pixel$^{-1}$. Rightmost (c) panel: Spectral index map obtained with the first six sub-band images of uSARA after smoothing with a common circular Gaussian beam of 5 arcsec. 
In \citet{askapdataset} are provided all sub-band images combined into the GIF \texttt{`SB8275-15\_uSARA'}, and {the spectral index maps of Abell 3395 obtained with uSARA and {\tt WSClean} in the GIF \texttt{`SpectralIndexMap\_Abell\_$3395$'}, together with a colour blind-friendly version in the GIF \texttt{`SpectralIndexMap\_Abell\_$3395$\_colorblind\_friendly'}.}
\label{A3391usara}}
\end{figure*}

\begin{table*}
\begin{tabular}{| *{9}{c|} }
\hline
\hline
\textbf{A3395 Phoenix} & $S_{887~{\rm MHz}}$  & $S_{923~{\rm MHz}}$  & $S_{959~{\rm MHz}}$ & $S_{995~{\rm MHz}}$  & $S_{1031~{\rm MHz}}$  & $S_{1067~{\rm MHz}}$  & $S_{1103~{\rm MHz}}$ & $S_{1139~{\rm MHz}}$  \\
    \hline
uSARA model   &   29.3 & 24.3 & 14.5 & 10.4 & 13.5 & 15.2 & 7.3 & 7.4  \\
    \hline
{\tt WSClean} restored image &  30.2 & 25.1 & 25.2 & 20.0 & 25.7 & 21.6 & 10.0 & 8.3 \\
\hline
{\tt WSClean} smoothed model  &  27.2 & 22.2 & 23.8 & 18.6 & 23.4 & 19.7 & 9.5 & 7.1  \\
\hline
\end{tabular}
\caption{Integrated flux density values in [mJy] of the diffuse phoenix source in Abell 3395 for each SPW imaged with uSARA and {\tt WSClean}.
\label{tab:fluxA3391}}
\end{table*}


\subsection{Imaging Settings}

To perform monochromatic sub-band imaging, we split the selected wide-band data into effective channels, or spectral windows (SPWs), across the full frequency band. More precisely, data from all three fields were binned uniformly into eight spectral windows with a bandwidth of 36 MHz each, under the assumption of nearly flat spectral behaviour within a spectral window. Data sizes per spectral window range from $\sim0.8$ to $\sim1.2$ GB. To further demonstrate the scalability of uSARA, we also imaged the third field (SB9442-35) over the full frequency band (covering 288 MHz) to form a single monochromatic image from $\sim7.5$ GB of data. We chose not to generate full-band monochromatic images for the other two fields (SB8275-15 and SB9351-12) since they host sources known to exhibit very steep spectral behaviour \citep[specifically, the galaxy clusters Abell 3391 and SPT-CL 2023-5535; ][]{2021A&A...647A...3B,2020ApJ...900..127H} and because both fields contain bright, large-scale artefacts.

According to the ASKAP Science Observation Guide\footnote{\url{https://confluence.csiro.au/display/askapsst/ASKAP+Survey+Science}}, recommended image size is set by the full width at half maximum (FWHM) of the primary beam. ASKAP beam pointings have a primary beam FoV approximated by a circular Gaussian with FWHM of $1.09\lambda_{\rm obs} / D$ \citep{2021PASA...38....9H}, where the observing wavelength, $\lambda_{\rm obs}$, is about 0.3 meters (at 1000 MHz) and $D$ is the dish diameter of a single ASKAP dish (12 m). We calculated the FWHM of a given beam pointing to be $\sim 1.56^{\circ}$ at the middle of the frequency band. For the first two selected ASKAP fields (SB8275-15 and SB9351-12), we noted the presence of bright sources lying just outside of the primary beam that created sidelobes and decided to image a FoV covering twice the FWHM. For the third FoV (SB9442-35), we did not find any bright sources directly outside of the primary FWHM and chose to image a FoV covering 1.6 times the FWHM. In both cases, the imaged FoV is well beyond the FWHM of the primary beam. With a maximum baseline of 6268~m, and a native instrumental resolution between 9 and 12 arcsec over the bandwidth, we selected a cell size of 2.2 arcsec pixel$^{-1}$, corresponding to a super-resolution factor of $2$ at the highest frequency. Under these considerations, the reconstructed images of fields SB8275-15 and SB9351-12 are $5500 \times 5500$ pixels in size, and those of the field SB9442-35 are $4096 \times 4096$ pixels. Unlike the output image of the CLEAN-based algorithm -- by design restricted to the instrumental resolution through the application of the restoring beam to its estimated non-physical model image -- the uSARA image retains super-resolution at the pixel size. On a final note, although the FWHM of the primary beam is used to determine the imaged FoVs, no primary beam correction is applied to the reconstructions in this work.

Systematic imaging was carried out on the three selected ASKAP fields with both uSARA and {\tt WSClean} using the same imaging settings where applicable. A summary of the imaging settings is listed in Table~\ref{tab: imgparams}. Our parallelised and automated imaging framework implemented in MATLAB, and the C++ software {\tt WSClean}, were both run on Cirrus\footnote{\url{http://www.cirrus.ac.uk}}, a UK Tier2 high-performance computing (HPC) service, comprising 280 compute nodes. Each node has 256~GB of memory and 36 CPU cores (with two hyperthreads each). Parameter choice in both algorithms is summarised in the following paragraphs.  

\paragraph*{{\tt{WSClean}} parameters.} 

The state-of-the-art {\tt WSClean} imager corrects for the $w$-effect via the $w$-stacking approach \citep{2014MNRAS.444..606O, 2017MNRAS.471..301O}. Monochromatic images of each of the selected ASKAP fields were made with Multi-scale CLEAN, using a gain factor of 0.8. Auto-masking was enabled down to 2.5 times the standard deviation of the {\tt WSClean} residual image, computed at the start of every major deconvolution iteration using the median absolute deviation estimator. The stopping criteria were set to a maximum iteration number of $10^6$ and a `cleaning' threshold equal to the estimated standard deviation of the residual image. Briggs robust weighting, with robust parameter $-0.25$ was considered in all the imaging experiments. These weights were stored in the {\tt IMAGING WEIGHT} column of the measurement set tables and were extracted and utilised in uSARA imaging. The number of $w$-stacks considered by {\tt WSClean} is set automatically based on the theoretical bound derived in \citet{2014MNRAS.444..606O} and the available compute resources (see Sec.~\ref{sec:comp-cost} for more details). For future reference, we note that CLEAN reconstructions are the so-called restored images, obtained as the sum of the non-physical model image convolved with the associated restoring beam, and the residual image. To support our flux density analysis, we also consider the {\tt WSClean} model images convolved with the restoring beam, referred to as smoothed model images.

\paragraph*{uSARA parameters.}\label{sec:usara_params} 

With the uSARA algorithm being implemented in MATLAB, data and associated information {(including the standard deviation of the data domain noise)} were extracted from the measurement set tables as a collection of MAT files using a dedicated Python script relying on the `Pyxis' library (standing for Python Extensions for Interferometry Scripting, as part of the {MeqTrees software package;}~\citealp{Noordam10}).  We recall that Briggs data weights generated by {\tt{WSClean}} were also considered in uSARA imaging. Concerning the uSARA measurement operator, the number of the $w$-stacks used in the different imaging experiments was set in an automated manner, via the planning step supported by our imaging framework (see Tables~\ref{tab:timeA3391}--\ref{tab:time9442}, for details). In the reconstruction of the sub-band images of all three fields, the operator $\bm\Phi^\dagger \bm\Phi$ was encoded via the underpinning sparse de-gridding matrices ($\Gb_{p}$). For the full-band monochromatic imaging experiment of the field SB9442-35, the resulting large data size triggered the dimensionality reduction feature. The operator $\bm\Phi^\dagger \bm\Phi$ was therefore encoded via its underpinning holographic matrices ($\Hb_{p}$), reducing the memory requirements to host it by nearly a factor 5. The regularisation parameter $\lambda$ was fixed to the heuristic value proposed in \eqref{eq:heuristic} for the full-band image of the field SB9442-35. However, some adjustment was found necessary to achieve a high reconstruction quality in terms of resolution and sensitivity for all sub-band images of the three selected fields, whereby $\lambda$ was set from 0.7 to 0.8 times the heuristic. Higher values of the regularisation parameter resulted in somewhat smoother reconstructions and non-recovery of some fainter point sources, whereas lower values led to images highly contaminated by calibration artefacts. The applied adjusting factor may be partly attributed to the imperfection of the DIE calibration and the lack of DDE calibration, affecting the accuracy of the measurement operator.  Nonetheless, it is generally consistent with the findings of the theoretical study of uSARA's heuristic in the context of simulated RI data \citep{Terris22}. Finally, the stopping criteria of uSARA were set to their default values, including a maximum of 10 re-weighted minimisation tasks, and a relative variation between the image iterates of $0.0005$.

\subsection{Quantitative Analysis}

Focusing on the target sources of the selected observations, we provide their flux measurements and in-band spectral index maps obtained by the two imaging algorithms. Estimated model images from uSARA are in units of [Jy pixel$^{-1}$], whereas {\tt WSClean} restored images are, by design, in units of [Jy beam$^{-1}$]. For the sake of comparison, we normalised {\tt WSClean} restored images by the area of the associated restoring Gaussian beam, denoted by $A_{\rm beam}$ and given by
\begin{equation}
 A_{\rm beam}  = \frac{\pi \times B_{\rm MAJ} \times B_{\rm MIN}}{4 \times \log 2 },
\end{equation}
where $B_{\rm MAJ}$ and $ B_{\rm MIN}$ are the respective major and minor axes of the restoring beam in pixel units (\emph{i.e.} normalised by the cell size of 2.2 arcsec pixel$^{-1}$).

Diffuse emission of particular interest in our study presents a complex morphology with edges often blended into the background noise, as seen in {\tt WSClean} maps. As is common practice, we measure the total flux of diffuse structure within manually generated regions which roughly follow $\sim2\sigma_{\rm meas}$ contours of the source, where $\sigma_{\rm meas}$ is the root-mean-square (rms) noise measured in a nearby region void of sources from the {\tt WSClean} restored map. Regions were hand-drawn in the visualisation software SAOImageDS9 \citep{ds903} to closely follow the contours of recovered signal in both the uSARA and {\tt WSClean} maps, such that the same region was used when measuring flux density of the same diffuse source. Flux density measurements from uSARA images are expected to be lower than those measured from the {\tt WSClean} restored images, due to the bias introduced from the {\tt WSClean} residual map. For a more accurate comparison of flux density, we also provide measurements from {\tt WSClean} smoothed model images. Note that error measurements on flux densities are not reported since the Early Science and Pilot Survey ASKAP observations are not yet validated against standard flux catalogues. All reported flux measurements and statistics were obtained using the SAOImageDS9 software.

Spectral index maps were created to showcase how sources of interest change over their morphology in electron energy distribution and consequently, in their spectral energy distribution. Firstly, the sub-band maps were smoothed via convolution with a 2D circular Gaussian kernel with axes of 5 arcsec for uSARA images and 20 arcsec for {\tt WSClean} images. The spectral index maps were then obtained by fitting a first-order polynomial to the function ${\rm log}(S_{\nu}) = -\alpha {\rm log}(\nu)$, where $S_{\nu}$ is the flux density for a given beam area at a given frequency $\nu$ and $\alpha > 0$ is the spectral index. Only the first six sub-bands are considered when generating these maps since the last two sub-bands were consistently found to recover less diffuse signal for primary targets of interest (possibly attributed to the steepness of their spectra).

\section{Results}\label{sec:results}

\begin{figure*}
\centering
\includegraphics[width=\textwidth]{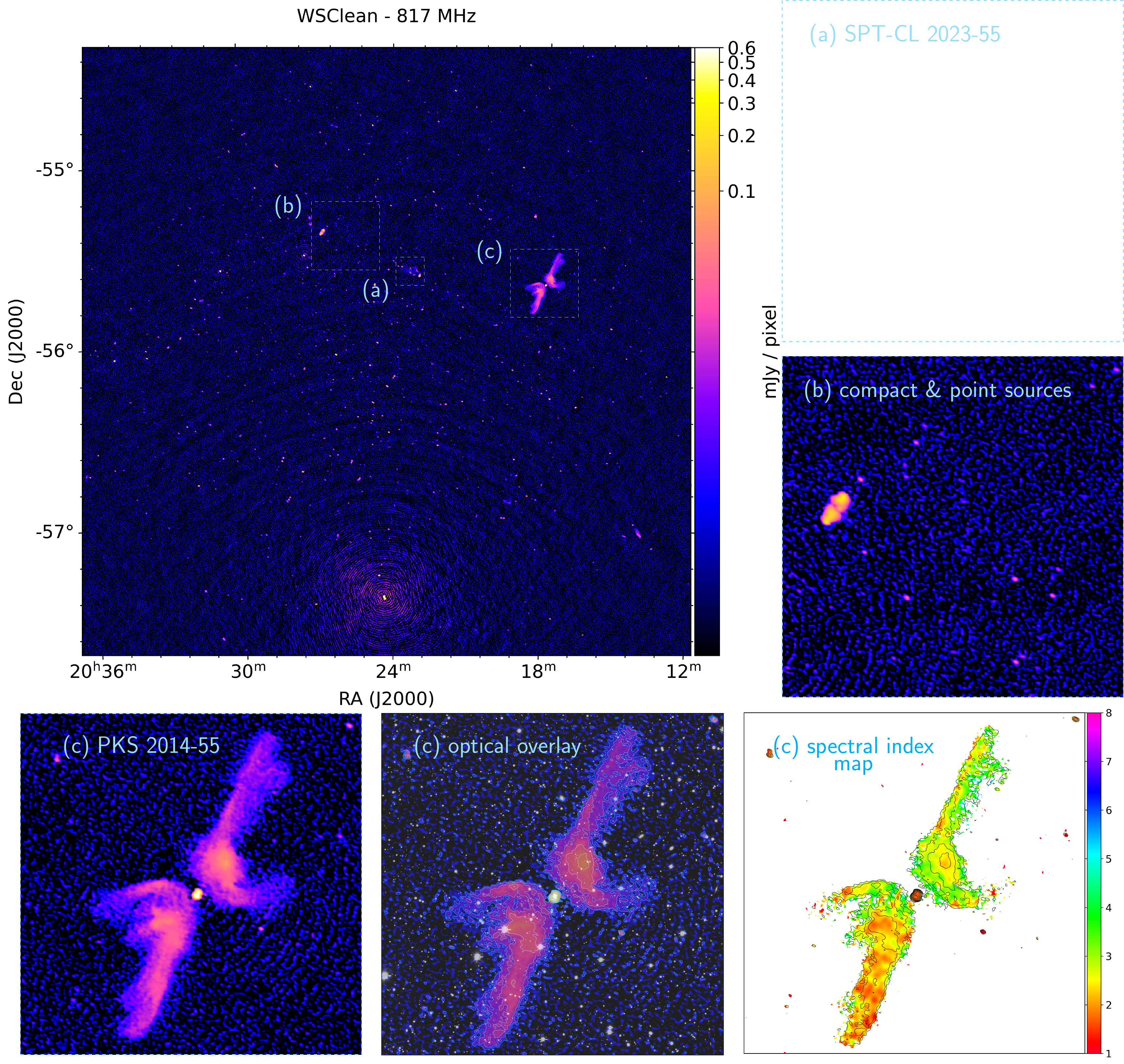}
\caption{SB9351-12 --{\tt WSClean}: Full FoV image covering the merging cluster SPT2023 and the X-shaped radio galaxy PKS 2014-55, at the first sub-band (SPW:1, centred at 817 MHz). This monochromatic image is a {\tt WSClean} restored image with a synthesised beam of $10.1 \times 16.4$ arcsec and rms noise of $\sigma_{\rm meas}\approx 60~\mu$Jy beam$^{-1}$ ($1.6~\mu$Jy pixel$^{-1}$). Panel (a) centred on the merging galaxy cluster SPT2023; panel (b) centred on a field containing compact and point sources; (c) panels centred on the X-shaped radio galaxy PKS 2014-55. Middle (c) panel: r-band optical image from DES overlaid with the {\tt WSClean} restored image, demarcated
by blue contours at the levels  $\{1.6 \times 2^n\}_{1 \leq n \leq 10} ~\mu$Jy pixel$^{-1}$. Rightmost (c) panel: spectral index map obtained with the first six sub-band images of {\tt WSClean} after smoothing with a common circular Gaussian beam of 20 arcsec. In \citet{askapdataset} are provided all sub-band images combined into the GIF \texttt{`SB9351-12\_WSClean'}, and {the spectral index maps of the X-shaped radio galaxy obtained with {\tt WSClean} and uSARA in the GIF \texttt{`SpectralIndexMap\_PKS\_2014\_55'}, together with a colour blind-friendly version in the GIF \texttt{`SpectralIndexMap\_PKS\_2014\_55\_colorblind\_friendly'}.}
\label{SPT2023wsclean}}
\end{figure*}

\begin{figure*}
\centering
\includegraphics[width=\textwidth]{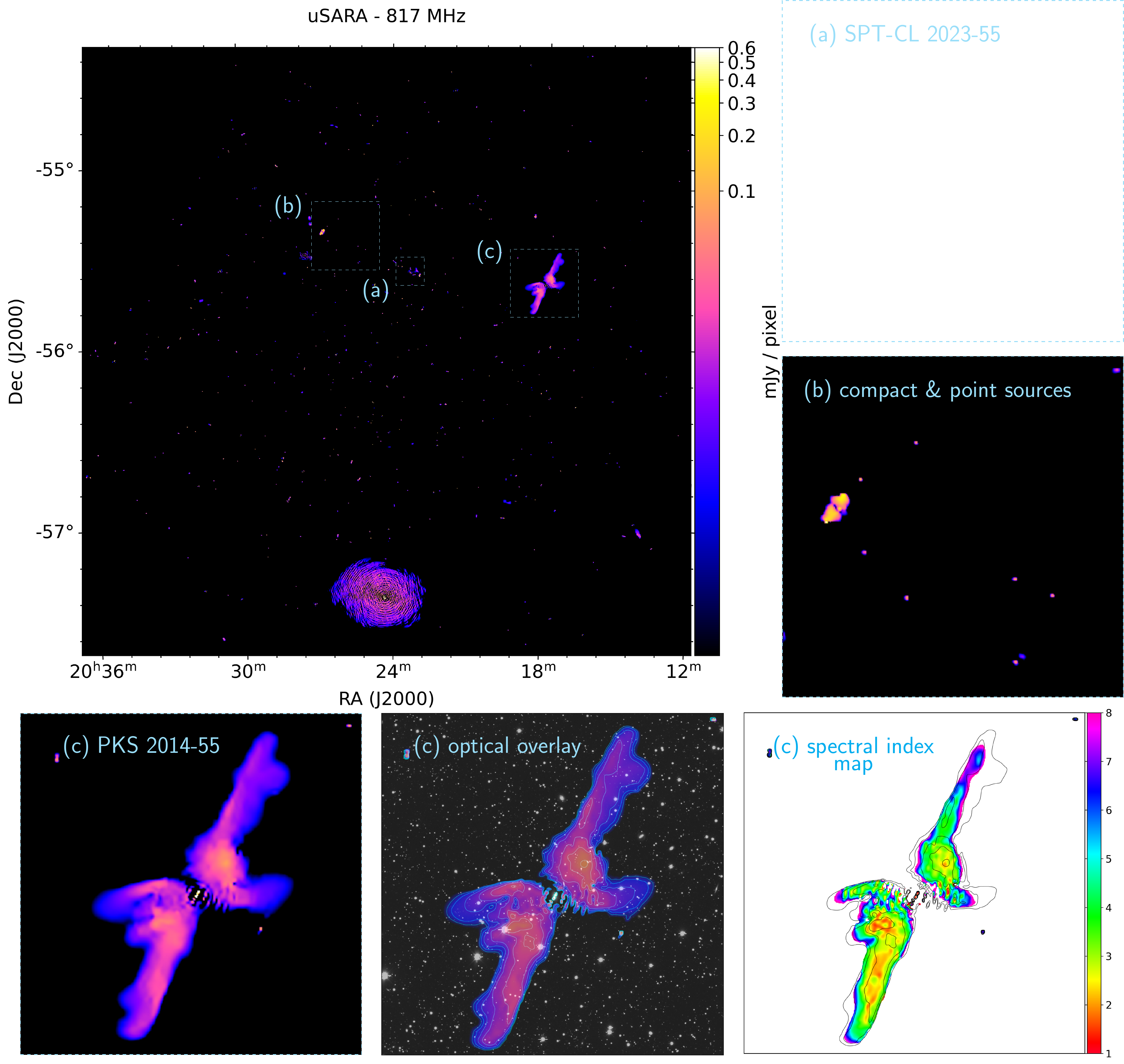}
\caption{SB9351-12 -- uSARA: Full FoV image covering the merging cluster SPT2023 and the X-shaped radio galaxy PKS 2014-55, at the first sub-band (SPW:1, centred at 817 MHz). This monochromatic image is a uSARA model with a pixel resolution of $2.2 \times 2.2$ arcsec. Panels are the same as described in Figure~\ref{SPT2023wsclean}. Middle (c) panel: r-band optical image from DES overlaid with the uSARA model image, demarcated by blue contours at the levels $\{2^n\}_{1 \leq n \leq 10} ~\mu$Jy pixel$^{-1}$. Rightmost (c) panel: spectral index map obtained with the first six sub-band images of uSARA after smoothing with a common circular Gaussian beam of 5 arcsec. In \citet{askapdataset} are provided all sub-band images combined into the GIF \texttt{`SB9351-12\_uSARA'}, and {the spectral index maps of the X-shaped radio galaxy obtained with uSARA and {\tt WSClean} in the GIF \texttt{`SpectralIndexMap\_PKS\_2014\_55'}, together with a colour blind-friendly version in the GIF \texttt{`SpectralIndexMap\_PKS\_2014\_55\_colorblind\_friendly'}.}
\label{SPT2023usara}}
\end{figure*}

In this section, we showcase high-resolution, high-fidelity images of our three selected fields produced by the uSARA algorithm and compare them to images made with multi-scale {\tt WSClean}. Select images are presented in Figures~\ref{A3391wsclean} -- \ref{9442usara}, showing the full imaged fields-of-view of our chosen ASKAP observations, and include zoomed-in views focusing on complex radio emission of interest and their associated optical images and spectral index maps. In \citet{askapdataset}, we provide FITS files of all spectral windows of the three selected fields imaged with both algorithms. For each field, images are also combined into animated GIF files to show how the recovered emission changes over the full frequency band. In what follows, we provide a detailed comparison of the morphology and flux density of specific sources between the uSARA and {\tt WSClean} images.

\subsection{First field: SB8275-15}

Beam 15 of SB8275 covers a FoV containing the complex merging galaxy cluster system Abell 3391 - Abell 3395. This field has been recently observed with the eROSITA X-ray space telescope where a warm gas bridge has been discovered between the cluster pair as part of a 15 Mpc intergalactic filament \citep{2021A&A...647A...2R}. The field also contains multiple bent-tail and Fanaroff-Riley class I and II (FR-I \& FR-II; \citealp{1974MNRAS.167P..31F}) radio galaxies, some belonging to the cluster system. A recent paper utilising mosaic images of SB8275 has confirmed more than 20 giant radio galaxies (at various redshifts) in the 30 deg$^2$ field \citep{2021A&A...647A...3B}.  

In Figures~\ref{A3391wsclean} \& ~\ref{A3391usara}, we present our images of the full FoV (3.36$^{\circ}$) of the first sub-band (SPW:1) of SB8275-15, imaged with {\tt WSClean} and uSARA, respectively. Both figures include zoomed-in views of the FR-I in Abell 3391 (a: top right panels), a FR-II cluster member in the east (b: middle right panels), and radio sources in Abell 3395 (c: bottom panels). The FR-I radio galaxies at the centre of Abell 3391 in the north and Abell 3395 in the south (see Table~\ref{tab: obs} for source names) are reconstructed with superior resolution by uSARA. This is most evident in the appearance of `braiding' and gaps in the plasma of the jets, which are not resolved in the {\tt WSClean} map. The FR-I in Abell 3391 is the brightest source in the field with a peak pixel flux of 20 mJy, as measured in the SPW:1 uSARA image, and 12 mJy as measured in the SPW:1 {\tt WSClean} image. Calibration errors in this field manifest as strong ring-like artefacts emanating from these bright FR-I radio galaxies. \citet{2021A&A...647A...3B} successfully carried out additional direction-dependent calibration to reduce the effect of these large-scale artefacts, however, we note that we performed no such additional calibration prior to imaging these data and therefore the extended radial artefacts remain in our final images. Over the full frequency band, source morphology and the structure of artefacts change per spectral window (see associated GIFs in \citealp{askapdataset}). 
 

\begin{table*}
\begin{tabular}{| *{9}{c|} }
    \hline
    \hline
\textbf{X-Shaped RG} & $S_{817~{\rm MHz}}$  & $S_{853~{\rm MHz}}$  & $S_{889~{\rm MHz}}$ & $S_{925~{\rm MHz}}$  & $S_{961~{\rm MHz}}$  & $S_{997~{\rm MHz}}$  & $S_{1033~{\rm MHz}}$ & $S_{1069~{\rm MHz}}$  \\
    \hline
uSARA model  & 758.8 & 680.3 & 591.7 & 526.7 & 454.0 & 335.2 & 286.8 & 268.2 \\
    \hline
{\tt WSClean} restored image &  756.4 & 721.4 & 629.3 & 546.2 & 511.8 & 409.5 & 369.0 & 352.1  \\
    \hline
{\tt WSClean} smoothed model &  743.7 & 707.1 & 613.8 & 532.7 & 498.9 & 399.7 & 357.5 & 308.4 \\
\hline
\end{tabular}
\caption{Integrated flux density values in [mJy] of the X-shaped radio galaxy PKS 2014-55 for each SPW imaged with uSARA and {\tt WSClean}.
\label{tab:fluxXshape}}
\end{table*}

\begin{table*}
\begin{tabular}{| *{9}{c|} }
    \hline
    \hline
\textbf{SPT2023 Relic} & $S_{817~{\rm MHz}}$  & $S_{853~{\rm MHz}}$  & $S_{889~{\rm MHz}}$ & $S_{925~{\rm MHz}}$  & $S_{961~{\rm MHz}}$  & $S_{997~{\rm MHz}}$  & $S_{1033~{\rm MHz}}$ & $S_{1069~{\rm MHz}}$  \\
    \hline
uSARA model   &  4.9 & 4.7 & 3.5 & 3.6 & 2.9 & 1.7 & 1.5 & 1.3 \\
    \hline
{\tt WSClean} restored image  &  5.7 & 5.8 & 4.3 & 4.1 & 3.8 & 3.5 & 3.0 & 3.4 \\
    \hline
{\tt WSClean} smoothed model  &  5.0 & 5.1 & 3.8 & 3.6 & 3.6 & 2.9 & 2.6 & 2.7 \\
\hline
\end{tabular}
\caption{Integrated flux density values in [mJy] of the radio relic in SPT2023 for each SPW imaged with uSARA and {\tt WSClean}. Central frequency of each SPW is listed in MHz. 
\label{tab:fluxSPT2023}}
\end{table*}

\subsubsection{Abell 3395}

The southern cluster Abell 3395 is made up of two sub-clusters, with separate X-ray peaks \citep{2021A&A...647A...2R}, indicating that this cluster is undergoing its own merger within the larger merging system. West of the central FR I in Abell 3395, there is a faint, diffuse source with a dim core connected by arms that extend to the north-west and south-west. As reported by \citet{2021A&A...647A...3B}, the peak intensity of this dim core does not clearly coincide with a host galaxy visible in optical maps, and therefore the source can not be classified as a typical radio galaxy associated with a host AGN. The diffuse source is possibly a so-called radio `phoenix' \citep[see][for classification]{2004rcfg.proc..335K}, re-ignited fossil plasma from past AGN activity which is no longer active. In the middle bottom panels of Figures~\ref{A3391wsclean} \& \ref{A3391usara}, {\tt WSClean} and uSARA images are overlaid as contours on an r-band optical image from DES DR1 \citep{2018ApJS..239...18A}. As seen in these optical overlays, a cluster member galaxy sits just south of the dim radio core, at RA, Dec 06h25m09.95s, -54$^{\circ}$30'34.4'', raising the possible scenario that old AGN emission from this galaxy has drifted or has been disturbed and shifted by hot gas in the cluster environment, leading to the observed faint emission. The structure of this phoenix candidate appears more clearly defined in our uSARA image, while the edges are much more blended into background noise in the {\tt WSClean} image. Most notably, the north-west and south-west limbs of the phoenix appear to pop out in the uSARA reconstruction, although they appear very faint and undefined in the {\tt WSClean} map.

We measure the flux density of this candidate phoenix source using identical polygonal regions hand-drawn to closely follow the total recovered signal in each of the  uSARA and {\tt WSClean} sub-band images, such that the same region is used to measure between the two imaging algorithms. In the {\tt WSClean} map of SPW:1, this polygonal region closely traces the $2\sigma_{\rm meas}$ contour line, where $\sigma_{\rm meas} = 2~\mu$Jy pixel$^{-1}$. Since the morphology of the phoenix source changes dramatically over the frequency band, a polygonal region was drawn for each spectral window. Interestingly, in the uSARA map of SPW:1 we find that the border of the phoenix's recovered signal can be traced by a $\sim 1~\mu$Jy pixel$^{-1}$ contour level (see middle bottom panel of Figure~\ref{A3391usara}), well below the estimated standard deviation of the noise in the image domain $\sigma$ (see Eq.~\ref{eq:sigma}) calculated as $6~\mu$Jy pixel$^{-1}$. This finding remains consistent through subsequent spectral windows, indicating that the uSARA algorithm successfully recovers real signal below the estimated noise levels. In the {\tt WSClean} maps, due to the blending of diffuse emission with background noise, the border of the phoenix is more clearly defined by contour lines at $2$ to $3\sigma_{\rm meas}$. 

For each sub-band, the measured flux densities from both uSARA and {\tt WSClean} images are listed in Table~\ref{tab:fluxA3391}. For the first two spectral windows, uSARA flux measurements of the candidate phoenix are greater; however, for all subsequent spectral windows, the {\tt WSClean} flux measurements are consistently greater. This is likely due to the fact that the measured flux density of this region in the {\tt WSClean} map is also integrated over the noise, which may increase for higher spectral windows and is amplified by artefacts emanating from the two bright FR I sources in the field. Indeed, lower flux densities are measured from the {\tt WSClean} smoothed model images, bringing them closer to uSARA values, particularly at the lower end of the frequency band. For such a faint source, we can see how the flux measurement from {\tt WSClean} restored images can be easily overestimated when mixed with a noisy background signal.

As apparent in Table~\ref{tab:fluxA3391}, the flux of the phoenix source reconstructed by uSARA drops off dramatically as the frequency increases, indicating a steep spectral index ($\alpha > 1$), as confirmed in \citet{2021A&A...647A...3B}. However, this fading over the frequency band is less dramatic in our {\tt WSClean} results, indicating that {\tt WSClean} may be possibly biased when wide-band imaging is deactivated. Comparing the spectral index maps of the phoenix (shown in the bottom right panels of Figures~\ref{A3391wsclean} \& \ref{A3391usara}), the general trend of steepening over the source morphology is similar. In the uSARA map, a steeper spectral index is seen in the phoenix's north-west limb. Interestingly, both spectral index maps show the phoenix hosting a steep core ($\alpha \sim 1.5$) surrounded by a halo of flatter emission ($\alpha < 1$), in contrast to the total intensity. This steep core is more clearly defined in the uSARA spectral index map. The fact that the brightest portion of the core is steeper in its spectral index than the surrounding fainter emission provides further evidence that this source may be an AGN remnant or phoenix. The emission around the potentially dormant core may exhibit a flatter spectral index because it has undergone gentle re-energisation \citep[\textit{e.g.}][]{2017SciA....3E1634D} from turbulence or small-scale shocks in the intracluster medium. Likewise, gentle re-energisation may explain why the ultra-steep emission in the north-west and south-west limbs is visible only at the lower end of the band (\textit{i.e.} subtle re-brightening of old AGN emission).

\subsection{Second field: SB9351-12}
Beam 12 of SB9351 covers a field containing the massive, merging galaxy cluster SPT-CL J2023-5535 (hereafter SPT2023) near the centre of the FoV and the X-shaped radio galaxy PKS 2014-55 on the western edge of the FoV. Since this beam observation lies on the western border of SB9351's full 30-square-degree field, we were unable to choose another beam observation that hosted the X-shaped radio galaxy closer to the pointing centre. Both the cluster and the radio galaxy have been recently studied: \citet{2020ApJ...900..127H} announced the discovery of a radio halo and radio relic in the merging cluster SPT2023 using the same EMU-PS observation we have selected, and \citet{2020MNRAS.495.1271C} used MeerKAT total intensity and polarisation observations to investigate the peculiar X-shaped morphology of PKS 2014-55. 

In Figures ~\ref{SPT2023wsclean} \& ~\ref{SPT2023usara}, we present our images of the full FoV (3.36$^{\circ}$) of the first sub-band (SPW:1) of SB9351-12, imaged with {\tt WSClean} and uSARA, respectively. Both figures include zoomed-in views of the galaxy cluster SPT2023 (a: top right panels), a field of compact and point sources (b: middle right panels), and the X-shaped radio galaxy (c: bottom panels). The bright quasar RX J2024.3-5723 at the southern edge of the pointing introduces radial ring-type artefacts, which propagate up to 1 deg in the field.  

In each of the zoomed-in views, uSARA shows higher resolution and more definition in the reconstruction of both compact and diffuse emission. However, the very faint emission of the radio halo in SPT2023 is not clearly recovered in the uSARA image. It is also apparent that some of the faintest point sources are missing from the uSARA image (see (b) panels of Figures~\ref{SPT2023wsclean} \& ~\ref{SPT2023usara}). This loss of the faintest point sources is likely attributed to the choice of the uSARA regularisation parameter. A lower value would enable the recovery of more of these point sources, but would also increase the amplitude of recovered calibration artefacts. 

\subsubsection{X-shaped Radio Galaxy}
As apparent when comparing panels (c) of Figures~\ref{SPT2023wsclean} \& ~\ref{SPT2023usara}, the X-shaped radio galaxy exhibits more clearly defined borders in our uSARA image. In the middle (c) panels of the same figures, {\tt WSClean} and uSARA emission of the X-shaped radio galaxy are overlaid as contours on an r-band optical image from DES DR1 \citep{2018ApJS..239...18A}. Again, we find that the border of the recovered uSARA signal of the X-shaped radio galaxy traces a contour level at $\sim 1~\mu$Jy pixel$^{-1}$, well below the estimated standard deviation of the image noise for this sub-band: $\sigma = 6~\mu$Jy pixel$^{-1}$. In contrast, the diffuse edges of the east and west wings of the X-shaped radio galaxy blend into background noise in the {\tt WSClean} map, such that the border is more clearly defined by $3\sigma_{\rm meas}$ contour lines, where $\sigma_{\rm meas} = 1.6~\mu$Jy pixel$^{-1}$. 

The total flux density of the radio galaxy is measured by summing the integrated flux density from three separate regions: the east wing, the core, and the west wing. The totalled flux density measurements from hand-drawn polygon regions for the lobes (roughly tracing emission bounded by the $2\sigma_{\rm meas}$ contour line in {\tt WSClean} images) and customised ellipse regions\footnote{Since the {\tt WSClean} image is convolved with the restoring beam, point sources are much more extended than they appear in our uSARA images. Therefore, we use differently sized ellipse regions to measure the flux density of the core of PKS 2014-55.} for the core are listed in Table~\ref{tab:fluxXshape}. The polygonal regions covering PKS 2014-55 were modified per each spectral window to more accurately follow the source morphology over the frequency band; however, identical regions were used to measure between the two imaging algorithms. As recorded in Table~\ref{tab:fluxSPT2023}, the flux density falls off as the frequency increases, indicating a steepening of the spectral index for this source. Except for the first spectral window, we see again that the flux is consistently greater in the {\tt WSClean} sub-band images, likely due to integration over the noise in the {\tt WSClean} map. When measuring from {\tt WSClean} smoothed model images, we find that the {\tt WSClean} flux densities decrease, bringing them more in line with uSARA measurements. 

Spectral index maps constructed from {\tt WSClean} and uSARA sub-band images are shown in the bottom right panels of Figures~\ref{SPT2023wsclean} \& ~\ref{SPT2023usara}. The general trend of steepening and flattening is consistent between the two maps, with more patches of flatter emission occurring in the lower portion of the east wing. This flattening is indicative of turbulent ``hot-spots'', coinciding with brightening seen in the intensity maps. Our uSARA spectral index map shows a dramatic steepening on the edges of the wings, but this is likely to be an artificial steepening since the diffuse structure at the edges is not recovered as well at higher frequencies (see associated GIF available in \citet{askapdataset}, demonstrating how the source structure changes with the frequency).

\begin{figure*}
\centering
\includegraphics[width=\textwidth]{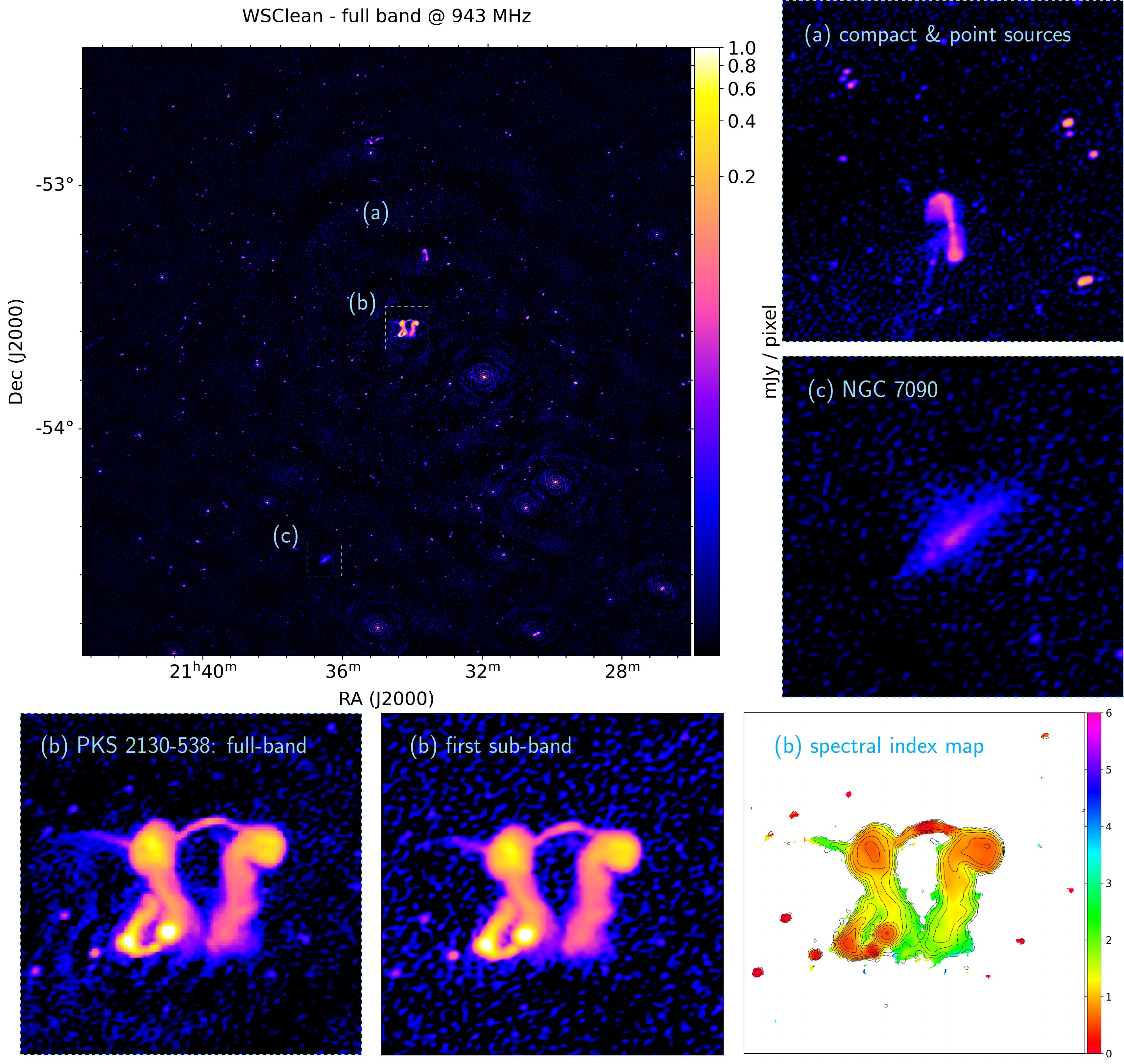}
\caption{SB9442-35 -- {\tt WSClean}: Full FoV image covering PKS 2130-538, formed using the full-band data (centred at 943 MHz). This monochromatic image is a {\tt WSClean} restored image with a synthesised beam of $11.8 \times 9.4$ arcsec and rms noise of $\sigma_{\rm meas}\approx 30~\mu$Jy beam$^{-1}$ ($1~\mu$Jy pixel$^{-1}$). Panel (a) centred on a field containing extended and point-like radio galaxies; panel (c) centred on the star-forming galaxy NGC 7090; (b) panels centred on ``the dancing ghosts'' (PKS 2130-538). Leftmost (b) panel: image made with the full-band data (centred at 943 MHz); middle (b) panel: image made with only the first sub-band of data (SPW:1, centred at 817), shown for a comparison of sensitivity; rightmost (b) panel: spectral index map made with the first six sub-band images of {\tt WSClean} after smoothing with a common circular Gaussian beam of 20 arcsec. In \citet{askapdataset} are provided all sub-band images combined into the GIF \texttt{`SB9442-35\_WSClean'}, and {the spectral index maps of ``the dancing ghosts'' obtained with {\tt WSClean} and uSARA in the GIF \texttt{`SpectralIndexMap\_PKS\_2130\_538'}, together with a colour blind-friendly version in the GIF \texttt{`SpectralIndexMap\_PKS\_2130\_538\_colorblind\_friendly'}.}
\label{9442wsclean}}
\end{figure*}

\begin{figure*}
\centering
\includegraphics[width=\textwidth]{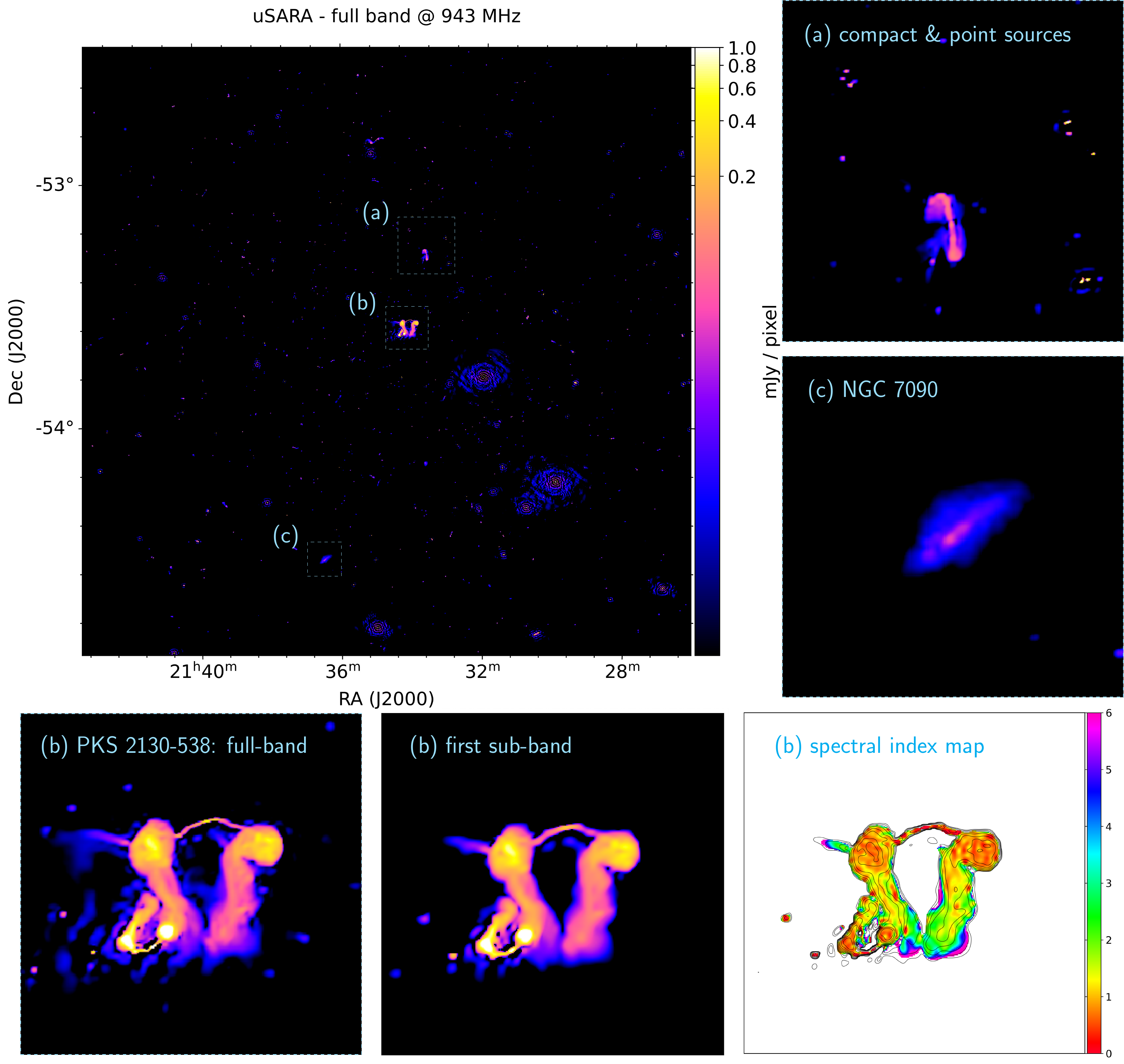}
\caption{SB9442-35 -- uSARA: Full FoV image covering PKS 2130-538, formed using the full-band data (centred at 943 MHz). This monochromatic image is a uSARA model with a pixel resolution of $2.2 \times2.2$ arcsec. Panels are the same as described in Figure~\ref{9442wsclean}. Rightmost (b) panel: spectral index map made with the first six sub-band images of uSARA after smoothing with a common circular Gaussian beam of 5 arcsec. In \citet{askapdataset} are provided all sub-band images combined into the GIF \texttt{`SB9442-35\_uSARA'}, and {the spectral index maps of ``the dancing ghosts'' obtained with uSARA and {\tt WSClean} in the GIF \texttt{`SpectralIndexMap\_PKS\_2130\_538'}, together with a colour blind-friendly version in the GIF \texttt{`SpectralIndexMap\_PKS\_2130\_538\_colorblind\_friendly'}.}
\label{9442usara}}
\end{figure*}

\begin{table*}
\begin{tabular}{| *{10}{c|} }
    \hline
    \hline
\textbf{Dancing Ghosts} & $S_{\rm fullband - 943~{\rm MHz}}$ & $S_{817~{\rm MHz}}$  & $S_{853~{\rm MHz}}$  & $S_{889~{\rm MHz}}$ & $S_{925~{\rm MHz}}$  & $S_{961~{\rm MHz}}$  & $S_{997~{\rm MHz}}$  & $S_{1033~{\rm MHz}}$ & $S_{1069~{\rm MHz}}$  \\
    \hline
uSARA model   &  117.2 & 131.0 & 126.0 & 121.2 & 116.3 & 112.1 & 107.7 & 104.6 & 101.8 \\
    \hline
{\tt WSClean} restored image & 115.5 & 128.8 & 124.5 & 120.0 & 115.3 & 111.2 & 107.2 & 104.2 & 101.0\\
    \hline
{\tt WSClean} smoothed model 
& 115.5 &  128.6 & 124.3 & 119.7 & 115.2 & 111.1 & 107.1 & 104.0 & 101.0 \\
\hline
\end{tabular}
\caption{Integrated flux density values in [mJy] of ``the dancing ghosts'' PKS 2130-53 for each SPW imaged with uSARA and {\tt WSClean}.
\label{tab:fluxPKS2130-53}}
\end{table*}


\subsubsection{SPT-CL J2023-5535}

As shown in panel (a) of Figure~\ref{SPT2023wsclean}, the radio halo in SPT2023 is barely recovered by {\tt WSClean}, appearing as a very faint increase in the noise across the inner regions of the cluster. The SPT2023 radio relic is apparent in the {\tt WSClean} map as a small, elongated arc at the western side of the cluster (we refer the reader to \citet{2020ApJ...900..127H} for a more detailed analysis of these cluster sources). Our uSARA image does not recover diffuse emission resembling a radio halo (see panel (a) of Figure~\ref{SPT2023usara}), likely due to the choice of regularisation parameter dampening signal below the estimated noise level; nonetheless, the radio relic appears more clearly defined and brighter than in the {\tt WSClean} image. We note that the classification of the halo and relic from \citet{2020ApJ...900..127H} was made using a full-band (288~MHz bandwidth) multi-frequency-synthesis image and that the signal in our individual sub-band images (36~MHz bandwidth) is much weaker. We report flux density measurements for the recovered relic source in SPT2023 in Table~\ref{tab:fluxSPT2023}. Flux density measurements of the relic from {\tt WSClean} smoothed model images are more in line with uSARA measurements, except for the last three sub-band where uSARA shows a decrease in flux. 

\subsection{Third field: SB9442-35}

Beam 35 of SB9442 is centred on the complex radio source PKS 2130-538. This peculiar source, nicknamed ``the dancing ghosts,'' is shaped by the jets and lobes of two radio galaxies in the Abell 3785 galaxy cluster. With the published catalogue and initial results of EMU-PS, \citet{2021PASA...38...46N} included an analysis of the morphology of PKS 2130-538. Two AGN hosts contribute to the observed emission: a radio galaxy in the north at the centre of a bright filamentary arch, and a second radio galaxy in the south at the centre of a smaller arch on the eastern lobe. Despite the advantage of ASKAP's resolution -- revealing previously unseen structure in PKS 2130-538 -- \citet{2021PASA...38...46N} point out that it is still unclear whether these two radio galaxies are superimposed or actually interacting with each other.  

Similarly to the previous two fields, eight sub-band images were reconstructed and used to obtain the flux density measurements, and the first six sub-band images were used to generate spectral index maps \citep[see associated GIF provided in][]{askapdataset}. We also formed a single full-band monochromatic image, for increased sensitivity, and to demonstrate the scalability of our imaging framework. In Figures ~\ref{9442wsclean} \& ~\ref{9442usara}, we present our monochromatic images of the full FoV (3.36$^{\circ}$) of SB9442-35, formed from the full-band data using {\tt WSClean} and uSARA, respectively. The figures also include zoomed-in views of a region of background sources (a: top right panels), the star-forming galaxy NGC 7090 (c: mid-right panels), and ``the dancing ghosts'' (b: bottom panels). Unlike the two previous fields, the SB9442-35 images do not exhibit large-amplitude calibration artefacts. In fact, only a few compact radio galaxies, catalogued by the Sydney University Molonglo Sky Survey (SUMSS; \citealp{sumss}), emanate localised ring-like artefacts, and do not hamper the recovery of our targets of interest.  

Overall, there is a clear difference between {\tt WSClean} and uSARA in terms of resolution. While uSARA recovers structure at higher resolution, diffuse components of the extended radio galaxy in panel (a) and the star-forming galaxy in panel (c) are also fully recovered when compared to the {\tt WSClean} map. However, the faintest point sources (near the noise level in the {\tt WSClean} image) in panel (a) are not fully recovered by uSARA. Again, we attribute this to the choice of the uSARA regularisation parameter.

\subsubsection{The Dancing Ghosts}

We focus on the complex emission of PKS 2130-538, displayed in panels (b) of Figures~\ref{9442wsclean} \& ~\ref{9442usara}. The left and middle panels represent a zoomed-in view of the ``dancing ghosts'' from the full-band image and the SPW:1 sub-band image, respectively. Looking at {\tt WSClean} images, as expected, the full-band restored image exhibits lower background noise and slightly more details in the source, in comparison to the sub-band map. As for uSARA images, a clear improvement in both resolution and sensitivity can be observed in the full-band image. The bridges, which consist of the jets from the northern and southern AGN, are more tightly collimated in the full-band image when compared to the sub-band image. Several faint point sources, missing from the sub-band image, also emerge. Interestingly, one can notice that the filamentary structure extending from the eastern lobe is more clearly defined and brighter in the uSARA reconstruction. This structure has a similar appearance to the synchrotron threads that were recently discovered in \citet{2020A&A...636L...1R,Dabbech22}. Recovery of this structure is an exciting result, as we may find more evidence of magnetic threads branching from and connecting extended lobes in radio galaxies with ultra-sensitive, super-resolved images. 

The bottom right panels of Figures~\ref{9442wsclean} \& ~\ref{9442usara} show spectral index maps of the dancing ghosts, inferred from the sub-band images. The uSARA spectral index map contains much more detailed information on the spectra of the turbulent lobes of the northern AGN. We also observe that the higher intensity hot-spots exhibit flatter spectral indices, with steepening as the emission traces southward. The second AGN at the centre of the south-east bridge shows a flat core, as expected, and a second set of jets that bend back toward the north-east lobe of the first AGN. In {\tt WSClean} maps, the second south-east lobe appears to blend back into the emission of the first north-east lobe, however, with uSARA we are able to see a distinct separation of these two portions, indicating that the emission may be somewhat superimposed when seen in projection. The steepness of the spectra in this region does not indicate a sort of re-energising from one lobe pushing onto the other, as it stays fairly consistent in a steep range between $1 < \alpha < 2$ in both spectral index maps. The north-east thread exhibits a sharp, collimated spectral trend from steep to slightly flat to steep again -- in contrast to the turbulent lobes -- with an index ranging from $2.9 \geq \alpha \geq 0.9$ and then $0.9 \leq \alpha \leq 3.2$ following the thread from west-to-east in our uSARA map. 

In Table~\ref{tab:fluxPKS2130-53}, flux densities for the Dancing Ghosts are reported. Integrated flux density was measured from identical polygonal regions tracing the full recovered signal in both uSARA and {\tt WSClean} maps, including the eastern and western lobes, the northern arch, the south-eastern jet, and the north-eastern filament. Unlike other diffuse sources of interest from previous fields, the uSARA flux densities are much more consistent with {\tt WSClean} flux densities. This presents an interesting case that may be explained by i) the source having overall flatter spectral behaviour, and ii) the lower noise level due to the absence of large-scale calibration artefacts in this field. At higher spectral windows, uSARA flux densities are even slightly greater than {\tt WSClean}, opposing the trend seen for the fainter, steeper diffuse sources of interest in previous fields. This may indicate that the faintest and steepest sources, with surface brightness near estimated noise levels, are more difficult to recover with uSARA, given our current choice of regularisation parameter. In both uSARA and {\tt WSClean} full-band images, the borders of the dancing ghosts are clearly defined by a contour level at $3~\mu$Jy pixel$^{-1}$, which is $3\sigma_{\rm meas}$ in the {\tt WSClean} full-band map (where $\sigma_{\rm meas} = 1~\mu$Jy pixel$^{-1}$). The uSARA imaging framework calculated the estimated standard deviation of the noise in the image domain as $\sigma = 8~\mu$Jy pixel$^{-1}$ for the full-band data, however, both uSARA and {\tt WSClean} images have captured signal below this value.

\section{Discussion}\label{sec:disc}

In this section, we discuss some specific points regarding the experiments performed with our novel automated and parallelised imaging framework, underpinned by the uSARA imaging algorithm \citep{Terris22,Dabbech22}. In comparison to the widely-used {\tt WSClean} imager, our uSARA-ASKAP images host greater detail at high resolution, resolving never-before-seen structure at scales up to 2.2 arcsec.

Arguably the most exciting feature of our reconstructed uSARA maps is that they successfully capture both compact and diffuse radio emission. Standard CLEAN imaging methods often require a compromise on resolution in order to gain sensitivity to diffuse emission, necessitating multiple maps at various resolutions to accurately recover emission on separate spatial scales. In many scientific publications and published radio surveys, imaging results are often found to be separated into one high-resolution map and another low-resolution map, where point sources and compact emission may be subtracted (usually through a crude technique that can leave holes in the image). Here, we have demonstrated that uSARA reconstructs images with both high resolution and sensitivity to diffuse emission, enabling advanced scientific analyses of physically complex sources. Thanks to uSARA's super-resolution and superior sensitivity to diffuse and extended components, we are also able to generate highly-detailed spectral index maps, which aid in the classification of our targeted sources.   

Moreover, we argue that uSARA can catapult the sensitivity and resolution of existing surveys. Unlike {\tt WSClean,} no residual image is added to the final uSARA reconstruction. It is therefore highly likely that the flux densities of low surface-brightness sources measured from uSARA images are a closer approximation to the true source flux.

\paragraph*{Calibration errors.} 
Several of the ASKAPsoft mosaic continuum images from early science and pilot fields show that calibrated data are still affected by radial artefacts that propagate -- up to several degrees in some cases -- from bright radio sources. For the ASKAP data considered in this work, the largest source of DDEs (manifested as antenna-based image modulations) is most likely attributed to ASKAP's synthesised beams. ASKAP's beamforming hardware applies digitised weights to the 188-elemental receivers and sums them to form a single primary beam at 1 MHz intervals. Currently, ASKAP uses a common algorithm to maximise signal-to-noise ratio to calculate the weights for beam synthesizing \citep[\emph{e.g.}][]{Jeffs2008,Ivashina2011}. Holographic observations of ASKAP's beam patterns \citep[see][for details]{Hotan16} show that their sensitivities vary over frequency from antenna to antenna. The complex-valued sensitivity pattern of the PAF beams therefore introduces DDEs which need to be modelled. Furthermore, antenna pointing errors can introduce direction-dependent antenna gains. ASKAPsoft corrects only for DIEs, and imperfections of the calibration undermine the accuracy of the measurement operator model. Consequently, reconstructed images can exhibit imaging artefacts and, more seriously, suffer from severely limited dynamic ranges. Examples of such artefacts can be seen in the field SB8275-15 containing the merging cluster system A3391-95, where large-amplitude ringing artefacts can be seen around the two bright FR-I radio galaxies (at the centres of the Abell clusters) in both uSARA and {\tt WSClean} reconstructions. In spite of the lack of DDE calibration, overall uSARA exhibits higher reconstruction quality than CLEAN. On a further note, uSARA can be easily plugged as the imaging module into a joint calibration and imaging framework \citep{Dabbech21}.

\paragraph*{In-band spectral index maps.}  
Spectral index maps obtained from sub-band monochromatic imaging with uSARA have shown to be more detailed than {\tt WSClean}. All sub-band images having a common cell size at least two times beyond the observation's nominal resolution at the highest sub-band, the spectral index maps were inferred using a small blurring kernel, preserving their high level of detail in comparison with {\tt WSClean} spectral index maps. 
We have found that the classification of some sources is more clearly defined based on the spectral behaviour exhibited in uSARA maps. Interestingly, some sources have shown steeper spectral indices in uSARA spectral index maps when compared to their {\tt WSClean} counterparts, which can be the result of the increase of sensitivity brought by uSARA. However, this spectral trend warrants further investigation by moving to wide-band deconvolution for a more precise spectral analysis.

\section{Computational performance}
\label{sec:comp-cost}
To assess the computational performance of our imaging framework, we report specific information for all uSARA imaging experiments in Tables~\ref{tab:timeA3391}--\ref{tab:time9442}. Details of the measurement operator are listed in these tables, including the number of processed visibilities $M$, the number of $w$-stacks $P$, the memory requirements to host its underlying sparse matrices ${{m}_{\Hb/\Gb}}$, and the computational cost in CPU core hour of $\bm{\Phi}^\dagger \bm{\Phi}$ pre-computation  C\textsubscript{$\bm{\Phi}^\dagger \bm{\Phi}$}. We also report the total compute time T\textsubscript{Image} and the computational cost in CPU core hour C\textsubscript{Image} of the deconvolution in the same tables. For comparison purposes, we report both T\textsubscript{Image} and C\textsubscript{Image} of {\tt WSClean} runs in Table~\ref{tab:time-wsclean}. For uSARA, the retained number of $w$-stacks in each experiment from the planning step determines the sparsity of the de-gridding/holographic matrices underpinning the measurement operator and consequently the memory requirements to host $\bm{\Phi}^\dagger \bm{\Phi}$. The decomposition of the data and associated measurement operator into smaller blocks, and the resulting number of deployed CPU cores are inferred from the data clustering step (see Section~\ref{ssec:prallel}). From Tables~\ref{tab:timeA3391}--\ref{tab:time9442}, one can notice that, in general, a larger number of visibilities will increase the computational cost of the measurement operator's pre-computation. Specific to FB iterations of uSARA, the compute time of the forward step is dominated by the Fourier transforms performed, while that of the backward step is driven by its sub-iterative nature. Both steps, being parallelised, are on par in terms of computing time, with the latter taking about 1.6 times longer on average. 

Although the number of $w$-stacks considered in {\tt WSClean} is significantly more important, the reported computational time and cost in CPU core hour of uSARA is about 20 times higher on average. The superior computational performance of the standard {\tt WSClean} RI imager is attributed to (i) its fast approximate data fidelity steps, whereby visibility gridding and de-gridding operations are only conducted few times, and (ii) its simplistic image model, in particular given the overall spatial compactness of the radio emission in the selected ASKAP fields, forcing multi-scale CLEAN to operate only on small-scales. However, the simple regularisation approach underpinning {\tt WSClean} comes at the expense of lower imaging quality. 

Even though {\tt WSClean} is about one order of magnitude faster than our imaging algorithm in its current MATLAB implementation prototype, substantial improvement of uSARA's computational performance is expected when migrated to a production implementation using C++ and parallel libraries.

\begin{table}
\begin{tabular}{cccccccc}
\hline
\hline
\textbf{SB8275} & $M$  & $P$ & $F$ &  ${{m}_{\Gb}}$  &   C\textsubscript{$\bm{\Phi}^\dagger \bm{\Phi}$} &  C\textsubscript{Image}& T\textsubscript{Image} \\
        & $\times 10^6$ &   &    & [GB] &[CPUh]&[CPUh]& [h] \\
     \hline
 \textbf{SPW:1}  & 56.8 & 9 & 42 &  153  & 92  & 268 & 5.5 \\
    \hline
\textbf{SPW:2}  & 75.8 & 10 & 49 &  186  & 101 & 313 & 5.4  \\
    \hline
\textbf{SPW:3}  & 74.5 & 9 & 56  &  216  & 93  & 378 & 5.7 \\ 
    \hline
\textbf{SPW:4}  & 76.9 & 9 & 64  &  228  & 112 & 400 & 5.9 \\ 
    \hline
\textbf{SPW:5}  & 76.0 & 9 & 64  &  241  & 112 & 422 & 6.1  \\
    \hline
\textbf{SPW:6}  & 76.9 & 8 & 64  &  277  & 113 & 481 & 6.2 \\
    \hline
\textbf{SPW:7}  & 69.2 & 9 & 64  &  239  & 115 & 397 & 5.8 \\
    \hline
\textbf{SPW:8}  & 66.3 & 9 & 64  &  244  & 103 & 467 & 6.9  \\
    \hline
 \end{tabular}
  \caption{SB8275-15 -- uSARA: settings and computational cost of the reconstruction of the sub-band images. $M$ is the number of measured visibilities; $P$ is the number of $w$-stacks; $F$ is the number of image facets used in the uSARA denoiser; $m_{\Gb}$ [GB] refers to the memory required to host the de-gridding matrices; C\textsubscript{$\bm{\Phi}^\dagger \bm{\Phi}$} [CPUh] is the computational cost of the pre-computation of the operator $\bm{\Phi}^\dagger \bm{\Phi}$ in CPU core hours; C\textsubscript{Image} [CPUh] is the computational cost of the deconvolution in CPU core hours; T\textsubscript{Image} is the time in hours of the deconvolution.  \label{tab:timeA3391}}
 \end{table}

\begin{table}
\begin{tabular}{cccccccc}
\hline
\hline    
\textbf{SB9351} &  $M$  & $P$ & $F$ & ${{m}_{\Gb}}$  &   C\textsubscript{$\bm{\Phi}^\dagger \bm{\Phi}$} &  C\textsubscript{Image}& T\textsubscript{Image} \\
        & $\times 10^6$ &   &    & [GB] &[CPUh]&[CPUh]& [h] \\
    \hline
\textbf{SPW:1}  & 60.8 & 7 &  36  & 140 & 55  & 358 & 8.1 \\ 
    \hline
\textbf{SPW:2}  & 50.3 & 6 &  36  & 141 & 54  & 367 & 8.3 \\ 
    \hline
\textbf{SPW:3}  & 43.7 & 6 &  36  & 131 & 55  & 354 & 8.4 \\
    \hline
\textbf{SPW:4}  & 49.8 & 7 &  36  & 134 & 58  & 342 & 8.1 \\ 
    \hline
\textbf{SPW:5}  & 60.7 & 9 &  36  & 131 & 67  & 349 & 8.1 \\ 
    \hline
\textbf{SPW:6}  & 53.2 & 8 &  36  & 137 & 69  & 364 & 8.5   \\ 
    \hline
\textbf{SPW:7}  & 63.0 & 9 &  42  & 147 & 78  & 388 & 8.2  \\ 
    \hline
\textbf{SPW:8}  & 50.6 & 8 &  36  & 139 & 72  & 379 & 8.8  \\ 
    \hline    
\end{tabular}
\caption{SB9351-12  -- uSARA: settings and computational cost of the reconstruction of the sub-band images. See the caption of Table~\ref{tab:timeA3391} for the complete list of the acronyms. \label{tab:timeSPT2023}}
\end{table}

\begin{table}
\begin{tabular}{cccccccc}
\hline
\hline    
\textbf{SB9442} &  $M$  & $P$ &  $F$ & ${{m}_{\Hb/\Gb}}$  &  C\textsubscript{$\bm{\Phi}^\dagger \bm{\Phi}$} &  C\textsubscript{Image}& T\textsubscript{Image} \\
          &$\times10^6$ &   &   &[GB]  &[CPUh]&[CPUh]&[h] \\
          \hline
\textbf{Full-band}& 467 & 19 & 64 & 80  & 281 & 770 & 4.3 \\ 
    \hline
\textbf{SPW:1}    & 50  & 6 & 36  & 94  & 40 & 210 & 5.2  \\ 
    \hline
\textbf{SPW:2}    & 62  & 7 &  36 & 107 & 50 & 221 & 5.5  \\  
    \hline
\textbf{SPW:3}    & 49  & 7 &  36 & 90  & 44 & 230 & 5.7  \\ 
    \hline
\textbf{SPW:4}    & 51  & 8 &  36 & 86  & 44 & 237 & 5.9  \\ 
    \hline
\textbf{SPW:5}    & 57  & 8 &  36 & 101 & 47 & 225 & 5.6  \\ 
    \hline
\textbf{SPW:6}    & 61  & 8 &  36 & 112 & 50 & 215 & 5.4  \\ 
    \hline
\textbf{SPW:7}    & 67  & 9 &  36 & 118 & 59 & 229 & 5.7 \\ 
    \hline
\textbf{SPW:8}    & 67  & 9 &  36 &  123 & 55 & 229 & 5.4  \\ 
    \hline
\end{tabular}
\caption{SB9442-35 -- uSARA: settings and computational cost of the reconstruction of sub-band and full-band images. See the caption of Table~\ref{tab:timeA3391} for the complete list of the acronyms. In the full-band imaging experiment, the operator $\bm{\Phi}^\dagger \bm{\Phi}$ is encoded via its underlying holographic matrices. We therefore report the memory occupied by these matrices, denoted by $m_{\Hb}$ [GB]. For all sub-band imaging experiments, we report the memory occupied by the de-gridding matrices denoted by $m_{\Gb}$ [GB]. \label{tab:time9442}}
\end{table}

\begin{table*}

\begin{tabular}{cccccccccc}
\hline
\hline  
& & \textbf{SB8275}   &  & &\textbf{SB9351} &    & &\textbf{SB9442}    &\\
\hline
& $P$ & C\textsubscript{Image}& T\textsubscript{Image}&$P$ &C\textsubscript{Image}&T\textsubscript{Image}&$P$&C\textsubscript{Image} & T\textsubscript{Image} \\
                & & [CPUh] & [h] & & [CPUh] & [h] & & [CPUh] & [h] \\
    \hline
\textbf{Full-band} &  -- & -- & -- & -- &   -- & -- &  72 & 58 & 0.8 \\ 
    \hline
\textbf{SPW:1}  & 89 & 22 & 0.6 & 72 & 22 & 0.6 & 72 & 14 & 0.4  \\
    \hline
\textbf{SPW:2} &  92 & 22 & 0.6 & 72 & 18 & 0.5 & 72 & 11 & 0.3  \\ 
    \hline
\textbf{SPW:3} &  96 & 22 & 0.6 & 75 & 14 & 0.4 & 72 & 11 & 0.3  \\ 
    \hline
\textbf{SPW:4} &  99 & 22 & 0.6 & 78 & 18 & 0.5 & 72 & 7 & 0.2  \\ 
    \hline
\textbf{SPW:5} & 103 & 22 & 0.6 & 81 & 18 & 0.5 & 72 & 11 & 0.3  \\ 
    \hline
\textbf{SPW:6} & 107 & 22 & 0.6 & 84 & 18 & 0.5 & 72 & 11 & 0.3  \\ 
    \hline
\textbf{SPW:7} & 110 & 22 & 0.6 & 87 & 22 & 0.6 & 72 & 11 & 0.3 \\ 
    \hline
\textbf{SPW:8} & 114 & 25 & 0.7 & 90 & 18 & 0.5 & 72 & 11 & 0.3  \\ 
    \hline
\end{tabular}
\caption{{\tt{WSClean}} settings and computational cost of all sub-band and full-band images: for each experiment, the number of $w$-stacks ($P$), the computational cost of the deconvolution in CPU core hours (C\textsubscript{Image}), and the imaging time in hours (T\textsubscript{Image}) are reported. Each experiment used a single node on Cirrus, leveraging its 36 CPU cores. Since each CPU core has two threads, 72 \textit{virtual} cores were detected, and consequently, the minimum number of $w$-stacks was set automatically to 72.\label{tab:time-wsclean}}
\end{table*}

\section{Conclusions}\label{sec:con}

This article presents a comprehensive study of a recently proposed monochromatic wide-field imaging framework for RI, validated on GB-sized, imperfectly calibrated data from Early Science and Pilot Survey ASKAP observations. The framework's underlying sparsity-based imaging algorithm, uSARA, yields image reconstructions with both super-resolution and high sensitivity to diffuse radio emission. Relying on a highly parallelised and automated implementation of the operators and functions involved, the imaging framework enables the formation of wide-field radio maps with large image dimensions. 

Focusing on science cases characterised by complex morphology exhibiting both compact and faint diffuse emission, we have carried out the validation of the uSARA imaging algorithm through flux density measurements and spectral index maps generated from sub-band monochromatic images, in comparison with the widely-used {\tt WSClean} imager. In spite of the large-scale artefacts due to imperfect calibration present in some of the selected fields, the uSARA images show better reconstruction quality overall in comparison to {\tt WSClean}, both in terms of resolution and sensitivity. Our uSARA-ASKAP images host more detailed structure of the targeted radio emission -- most clearly seen in the super-resolved jets and lobes of radio galaxies in each of the selected FoVs. In addition to high-resolution structure, faint diffuse emission has also been captured by uSARA, revealing more extended emission of intracluster radio sources which appeared blended into background noise in sub-band {\tt WSClean} maps.

An advantageous result of our super-resolved uSARA-ASKAP images is the ability to generate more detailed spectral index maps. High-resolution structure, resembling turbulent emission in the radio lobes of several imaged radio galaxies, appears to closely trace small changes in the steepness of the observed spectra. Furthermore, each of our primary sources of interest exhibit steeper spectra in the uSARA spectral index maps, attributed to the increase in sensitivity and resolution delivered by the algorithm. Nonetheless, our spectral analysis of the target sources remains preliminary and warrants a deeper study using wide-band imaging. Planned upgrades to the uSARA framework -- which will incorporate joint DDE calibration \citep{repetti17} and wide-band deconvolution \citep{thouvenin22a} -- guarantee the robustness of future images, and consequently, more precise spectral information across all frequency channels.

In terms of scalability of the proposed imaging framework, we have demonstrated that its fully automated and parallel measurement operators enable image reconstruction of data up to 7.5~GB in size. 
Larger data dimensions and fields-of-view   necessitate distributing the data and associated measurement operator into more blocks, and decomposing the image into more facets for the parallel application of the sparsity dictionary and its adjoint in the uSARA denoiser. By adding more computational resources, the time to solution can be maintained at a reasonable scale.
In its current MATLAB implementation, the computational cost of our imaging framework remains higher than the benchmark imager {\tt WSClean}. However, its migration to C++ leveraging parallel libraries, can substantially boost its computational efficiency, thus narrowing the computational gap with the state-of-the-art imager.

In the sequel to this series, Part II: ``AIRI validated on ASKAP data,'' we investigate uSARA's sister algorithm AIRI, the second deconvolution algorithm built into our parallel automated imaging framework. AIRI differs from uSARA by exploiting learned Deep Neural Network (DNN) denoisers in lieu of uSARA's proximal operator in the underpinning FB algorithmic structure \eqref{eq:fb}. AIRI was already very recently and briefly demonstrated on MeerKAT data \citep{Dabbech22}. The algorithm will be validated on the same challenging ASKAP data as here, with the aim to further demonstrate its potential to deliver further imaging precision and faster reconstruction than uSARA.

\section*{Acknowledgements}
The first two authors contributed equally to this work. This work was supported by the UK Research and Innovation under the EPSRC grants EP/T028270/1 and EP/T028351/1, and the STFC grant ST/W000970/1. The research used Cirrus, a UK National Tier-2 HPC Service at EPCC funded by the University of Edinburgh and EPSRC (EP/P020267/1). ASKAP, from which the data under scrutiny originate, is part of the Australia Telescope National Facility managed by CSIRO. This project used public archival data from the Dark Energy Survey (DES).

\section*{Data Availability}
The ASKAP data underlying this article (calibrated visibilities and mosaic images of Scheduling Blocks) are made publicly available for viewing and download on the \href{https://data.csiro.au/collections/#domain/casdaObservation/search/}{CSIRO ASKAP Science Data Archive} (CASDA; \citealp{2017ASPC..512...73C}), and can be accessed with the unique Project Identifiers AS034 and AS101. The reconstructed images in FITS format as well as the GIF files showing the imaged fields over the spectral windows are made available in \citet{askapdataset}. The uSARA and AIRI code will become available in a later release of the Puri-Psi library for RI imaging.



\bibliographystyle{mnras}
\bibliography{ASKAP-uSARA} 





\bsp	
\label{lastpage}
\end{document}